\documentclass[onecolumn]{IEEEtran}
\usepackage[final]{graphicx}
\usepackage{cite}
\usepackage{amssymb}
\usepackage{amsmath}
\usepackage{amsthm}
\usepackage{amsfonts}
\usepackage{bm}
\usepackage{makecell}
\usepackage{float}
\usepackage{booktabs}
\usepackage{multirow}
\usepackage{subfigure}
\usepackage{subfigmat}

\title{Robust SINR-Constrained Symbol-Level Multiuser Precoding with Imperfect Channel Knowledge}

\author{Alireza~Haqiqatnejad,~\emph{Student Member,~IEEE},~Farbod~Kayhan,~and Bj\"{o}rn~Ottersten,~\emph{Fellow,~IEEE}
        \thanks{The authors are with the Interdisciplinary Centre for Security, Reliability and Trust (SnT), University of Luxembourg, Luxembourg City L-1855, Luxembourg (e-mail:~alireza.haqiqatnejad@uni.lu; farbod.kayhan@uni.lu; bjorn.ottersten@uni.lu).}
        \thanks{This paper has been presented in part at the IEEE Global Communications Conference (GLOBECOM), Abu Dhabi, UAE, 2018.}
}

\newtheorem{theorem}{Theorem}

\newtheorem{lemma}[theorem]{Lemma}

\newcommand{\Del}{\mathrm{\pmb{\Delta}}}

\newcommand{\HHH} {\mathrm{\pmb{H}}}
\newcommand{\QQQ} {\mathrm{\pmb{Q}}}
\newcommand{\BBB} {\mathrm{\pmb{B}}}
\newcommand{\bbb}{\mathrm{\pmb{b}}}
\newcommand{\uuu}{\mathrm{\pmb{u}}}

\newcommand{\EEE}{\mathrm{\pmb{E}}}
\newcommand{\WWW}{\mathrm{\pmb{W}}}
\newcommand{\ZZZ}{\mathrm{\pmb{Z}}}
\newcommand{\I}{\mathrm{\pmb{I}}}
\newcommand{\J}{\mathrm{\pmb{J}}}
\newcommand{\A}{\mathrm{\pmb{A}}}
\newcommand{\DDD}{\mathrm{\pmb{\Sigma}}}
\newcommand{\FFF}{\mathrm{\pmb{F}}}

\newcommand{\REAL}{\mathrm{Re}}
\newcommand{\IMAG}{\mathrm{Im}}

\newcommand{\YYY}{\mathrm{\pmb{Y}}}
\newcommand{\CCC}{\mathrm{\pmb{C}}}
\newcommand{\XXX}{\mathrm{\pmb{X}}}

\newcommand{\GGG}{\mathrm{\pmb{G}}}
\newcommand{\OOO}{\mathrm{\pmb{0}}}
\newcommand{\aaa}{\mathrm{\pmb{a}}}
\newcommand{\h}{\mathrm{\pmb{h}}}
\newcommand{\qqq}{\mathrm{\pmb{q}}}
\newcommand{\eee}{\mathrm{\pmb{e}}}
\newcommand{\s}{\mathrm{\pmb{s}}}

\newcommand{\fff}{\mathrm{\pmb{f}}}

\newcommand{\ccc}{\mathrm{\pmb{c}}}
\newcommand{\x}{\mathrm{\pmb{x}}}
\newcommand{\yyy}{\mathrm{\pmb{y}}}

\newcommand{\www}{\mathrm{\pmb{w}}}

\newcommand{\EXP}{\mathbb{E}}
\newcommand{\PR}{\mathbb{P}}
\newcommand{\TR}{\mathrm{Tr}}
\newcommand{\VEC}{\mathrm{vec}}

\newcommand{\INF}{\mathrm{inf}}

\newcommand{\TT}{\mathrm{T}}
\newcommand{\C}{\mathbb{C}}

\newcommand{\diag}{\mathop{\mathrm{diag}}}

\begin{document}

\maketitle

\begin{abstract}
In this paper, we address robust design of symbol-level precoding for the downlink of multiuser multiple-input multiple-output wireless channels, in the presence of imperfect channel state information (CSI) at the transmitter. In particular, we consider two common uncertainty models for the CSI imperfection, namely, spherical (bounded) and stochastic (Gaussian). Our design objective is to minimize the total (per-symbol) transmission power subject to constructive interference (CI) constraints as well as users' quality-of-service requirements in terms of signal-to-interference-plus-noise ratio. Assuming bounded channel uncertainties, we obtain a convex CI constraint based on the worst-case robust analysis, whereas in the case of Gaussian uncertainties, we define probabilistic CI constraints in order to achieve robustness to statistically-known CSI errors. Since the probabilistic constraints of actual interest are difficult to handle, we resort to their convex approximations, yielding tractable (deterministic) robust constraints. Three convex approximations are developed based on different robust conservatism approaches, among which one is introduced as a benchmark for comparison. We show that each of our proposed approximations is tighter than the other under specific robustness conditions, while both always outperform the benchmark. Using the developed CI constraints, we formulate the robust precoding optimization as a convex conic quadratic program. Extensive simulation results are provided to validate our analytic discussions and to make comparisons with existing robust precoding schemes. We also show that the robust design  increases the computational complexity by an order of the number of users in the large system limit, compared to its non-robust counterpart.
        
\end{abstract}


\section{Introduction}
\IEEEPARstart{M}{ultiuser} precoding is a well-known technique to enhance the achievable throughput and the reliability of communication in a downlink multiuser multiple-input multiple-output (MU-MIMO) wireless system. In principle, this improvement is brought by employing multiple antennas at the transmitter, which enables more degrees of freedom to manage the channel-induced multiuser interference (MUI). In most applications, however, the system may be subject to some crucial system-centric and/or user-specific requirements, e.g., total/per-antenna power budget or quality-of-service (QoS) targets. In such scenarios, the precoding design problem needs to be constrained by the given requirements while aiming at optimizing a certain objective function; this kind of design is often called objective-oriented precoding optimization \cite{tb_convex}. Among a variety of design criteria, a frequently addressed one is the QoS-constrained power minimization; see e.g. \cite{tb_vis,tb_sinr,tb_sol_str}.

In general, multiuser precoding schemes can be categorized in two groups, namely, conventional (block-level) techniques and symbol-level techniques. In the conventional precoding, the precoder typically exploits the channel knowledge in order to suppress/eliminate the MUI, regardless of the current users' symbols \cite{tb_opt,vec_per}. On the contrary, in the symbol-level design, the basic idea is to convert the (potential) MUI into a desired received signal component, i.e., into the so-called constructive interference (CI), by means of processing the transmit signal on a symbol-level basis \cite{tb_green, slp_con}.

In reality, assuming perfect channel state information (CSI), either statistically or instantaneously, is rather impractical due to various inevitable channel impairments such as imperfect channel estimation, limited feedback, or latency-related errors \cite{feedback,unc_sens,unc_imp}. However, potential performance improvements may no longer be offered by multiuser precoding if accurate CSI is not available at the transmitter, broadly because precoding techniques are quite sensitive to channel uncertainties \cite{unc_sens}. One may expect an even more adverse effect of imperfect channel knowledge on the symbol-level precoder's performance, due to the fact that the promised efficiency (extremely) depends on the satisfaction of CI constraints in order to successfully accommodate each (noise-free) received signal in the proper CI region. To alleviate this reliance, the problem of designing a multiuser precoder that is robust to channel uncertainties becomes of practical interest.


The channel uncertainty region is commonly considered to be either ellipsoidal or stochastically-distributed, or a combination of both, e.g., see \cite{power_allo}.
Under the ellipsoidal uncertainty model, usually no assumption is made on the distribution of the CSI error, but rather the error is supposed to always lie within a norm-bounded  region. When the Frobenius/Euclidean norm is adopted, the model is sometimes called spherical uncertainty \cite{several_palo}. This kind of modeling, which ultimately leads to a worst-case analysis, is known to appropriately capture the bounded uncertainties resulted from quantization errors \cite{maximin}.
The stochastic uncertainty model, on the other hand, assumes statistical properties for the CSI error.
In scenarios with channel estimation at the transmitter side, such modeling is particularly suitable since the error in the estimation process can often be treated as a Gaussian random variable \cite{estimation_poor}.

With a particular focus on MU-MIMO broadcast channels, a wide variety of robust schemes can be found in the literature on conventional multiuser precoding, addressing both spherical and stochastic uncertainty models. In this context, most of the existing research considers either of the QoS-constrained power minimization or the max-min fairness with power constraints as the design formulation. Under norm-bounded CSI uncertainty, the QoS problem is typically constrained by the worst case of users' signal-to-interference-plus-noise ratio (SINR), resulting in highly conservative design approaches; see, for example, \cite{unc_partial,unc_conic,unc_eigen} as some notable research in this direction. These worst-case SINR requirements can also be translated to worst-case minimum mean-square error (MMSE) constraints \cite{unc_tract,tb_qos}. With the assumption of (normally-distributed) stochastic CSI error, the QoS targets are usually expressed by probabilistic SINR constraints as in \cite{unc_sta,chance_holger,davidson_prob}, or in terms of equivalent rate-outage probability restrictions \cite{outage_spec,outage_cri,outage_conic}. Given in either form, the stochastically-robust schemes mostly apply the robust (chance-constrained) optimization techniques introduced in \cite{safe} and \cite{bertsimas}.

The robust design of the symbol-level precoding is not well investigated in the literature. A worst-case robust analysis is provided in \cite{slp_chr} to design the symbol-level precoder with norm-bounded CSI errors, addressing the power minimization and the max-min fairness problems. It is, however, important to notice that as far as the symbol-level power minimization problem is concerned, the norm-bounded uncertainty model might not yield an efficient solution. This modeling ultimately leads to a worst-case conservatism which inherently increases the transmission power, though enhancing the users' symbol error probability. To the best of the authors' knowledge, there is no published work to date with the aim of developing a stochastically-robust symbol-level design formulation. It is worth mentioning that a precoding optimization with outage probability constraints based on a symbol-level approach is presented in \cite{unc_chr}, however, the goal is to achieve robustness to noise uncertainty, but not to any type of channel uncertainties.

In this paper, we study the problem of symbol-level precoding design in the presence of channel uncertainty. Our goal is to optimize the (total) transmission power under joint CI and SINR constraints. In the optimization problem, the CI constraints are formulated by adopting the distance-preserving constructive interference regions (DPCIR), introduced in \cite{slp_gen}. We consider both spherical and stochastic uncertainty models. In order to obtain a robust formulation for the original CI constraint, it is essential to characterize the uncertain component appearing in the CI inequality as a result of the imperfect CSI. Our primary challenge, however, is to obtain tractable convex approximations for the resulting robust formulation, ensuring that the desired constraint is met for any realization of the CSI error within the uncertainty set. The relative tightness of the derived approximations, which (roughly speaking) measures the cost of tractability, then becomes of interest. Having the convex robust constraints, the subsequent modification of the precoding design problem is straightforward due to the fact that the only part of the problem being affected by the channel uncertainty is the CI constraint. However, the complexity of the robust precoding optimization might be different from the original problem. Accordingly, the main contributions of this paper are listed as below:
\begin{itemize}
        \item[1.] We propose some modifications to the CI constraints according to both bounded and stochastic (Gaussian) uncertainty models. In the scenario with norm-bounded CSI uncertainty, we obtain a robust second-order cone constraint based on the notion of worst-case robust analysis. For Gaussian CSI errors, we redefine the CI constraint as a chance-constrained inequality for which we develop two approximate convex robust alternatives based on the probability bounding idea and the safe approximation method. Both the approximations are expressed as convex second-order cone constraints, hence are efficiently computable. We further obtain a third robust reformulation based on the well-known idea of sphere bounding as our benchmark for comparison. Under a specific condition, we show that the safe convex approximation can also be expressed as a convex second-order cone constraint. This allows us to compare the relative tightness of the obtained robust approximations through analytic discussions, which will be validated using simulation results. Our results indicate that both the proposed robust schemes provide tighter approximations than that obtained from the sphere bounding method.
        
        \item[2.] We cast the robust QoS-constrained (symbol-level) power optimization as a convex conic quadratic program (CQP) for both uncertainty models and all the proposed robust formulations of the CI constraint. We then analyze and compare the complexities of robust and non-robust precoding design problems, through which we indicate that either of the proposed robust approaches leads to a higher computational complexity compared to that of the non-robust problem, by a dominating order of the number of users as the system dimension grows to infinity.
\end{itemize}

\noindent{\bf{Organization:}} The rest of this paper is organized as follows. We describe the system and uncertainty models in Section \ref{sec:sys}. In Section \ref{sec:slp}, first we briefly explain the original (non-robust) CI constraints in the symbol-level precoding problem. We then define robust counterparts for the desired CI inequalities and develop reformulations in the form of approximate convex restrictions. We also provide analytic discussions on the tightness of approximation in this section. In Section \ref{sec:optprob}, we cast the robust symbol-level precoding optimization problem and analyze the resulting computational complexity. Our simulation results are provided in Section \ref{sec:sim}. Finally, we conclude the paper in Section \ref{sec:con}.

\noindent{\bf{Notations:}} We use uppercase and lowercase bold-faced letters to denote matrices and vectors, respectively. The sets of real and complex numbers are represented by $\mathbb{R}$ and $\mathbb{C}$. For a complex input, $\REAL\{\cdot\}$ and $\IMAG\{\cdot\}$ respectively denote real and imaginary parts. For matrices and vectors, $  [\,\cdot\,]^T$  denotes transpose. For a (square) matrix $\A$, $|\A|$ and $\mathrm{Tr}(\A)$ respectively denote the determinant and the trace of $\A$, $\VEC(\A)$ stands for the vector obtained by stacking the columns of $\A$, and $\A\succeq 0$ (or $\A\preceq 0$) means that $\A$ is positive semidefinite (or negative semidefinite). For two square matrices $\A$ and $\BBB$ with identical dimensions, $\A\succeq\BBB$ means $\A-\BBB$ is positive semidefinite. Given two vectors $\x\in\mathbb{R}^n$ and $\yyy\in\mathbb{R}^n$, $\x\geq\yyy$ (or $\x\ngeq\yyy$) denotes the entrywise inequality. $\|\cdot\|_2$ and $\|\cdot\|_F$ represent the vector Euclidean norm and the matrix Frobenius norm, respectively. $\pmb{I}$, $\OOO$ and $\pmb{1}$ respectively stand for the identity matrix, the zero matrix (or the zero vector, depending on the context) and the all-one vector of appropriate dimension. The probability function and the statistical expectation are respectively denoted by $\PR\{\cdot\}$ and $\EXP\{\cdot\}$. The operators $\otimes$ and $\circ$ stand for the Kronecker and the Hadamard products, respectively.



\section{System and Uncertainty Model}\label{sec:sys}

We consider an MU-MIMO wireless broadcast channel in which a common transmitter (e.g., a base station), equipped with $N$ antennas, serves $K$ single-antenna users by sending independent data streams, where $K\leq N$. We denote by the row vector $\h_k\in\C^{1\times N}, k=1,...,K,$ the instantaneous (frequency-flat) fading channel of the $k$th transmit/receive antenna pair. In the downlink transmission, at any symbol instant $t=0,1,2,...$, independent data symbols $s_k(t),k=1,...,K,$ are to be conveyed to the users, with $s_k(t)$ denoting the intended symbol for the $k$th user. To simplify the notation, we focus on a specific symbol time and drop the time index $t$ throughout the paper. Each symbol $s_k$ is drawn from a finite equiprobable constellation set with unit average power, where all the constellation points have unbounded (Voronoi) decision regions.
We further assume, without loss of generality, that all the users employ identical $M$-ary modulation schemes.

We collect the desired symbols of all $K$ users in a vector denoted by $\s=[s_1,\ldots,s_K]^T\in\C^{K\times1}$. The symbol vector $\s$ is then mapped to $N$ transmit antennas yielding the transmit vector $\uuu=[u_1,\ldots,u_N]^T\in\C^{N\times1}$. This mapping is done with the use of an appropriately designed multiuser precoding module. In this paper, we adopt a symbol-level precoding (SLP) scheme based on a particular type of constructive interference regions, which will be discussed in more detail later. It is worth noting that unlike conventional (block-level) precoders, e.g., (regularized) zero-forcing or minimum mean square error, in symbol-level mapping there might be no explicit precoding matrix in general (relating the symbol vector $\s$ to the transmit vector $\uuu$) to be optimized. Instead, the optimal transmit signal $\uuu$ is obtained as a result of an objective-oriented precoding design on a symbol-level basis. At the receiver of the $k$th user, the observed signal is

\begin{equation}\label{eq:sys}
r_k = \h_k\uuu+z_k, \; k=1,...,K,
\end{equation}
where $z_k$ represents the additive circularly symmetric complex Gaussian noise distributed as $z_k\sim\mathcal{CN}(0,\sigma_k^2)$. The $k$-th user may use the conventional single-user detector based on the maximum-likelihood (ML) decision rule to optimally detect its desired symbol $s_k$, i.e., the structure of the receiver is independent of the precoder design. 

While it is assumed that all the users have perfect knowledge of their own channels, the transmitter normally has inaccurate CSI due to several reasons  such as imperfect channel estimation, limited (or delayed) feedback and quantization errors. By adopting a perturbation-based uncertainty model, the actual channel of user $k$ is expressed as

\begin{equation}\label{eq:H}
\h_k = \hat{\h}_k + \eee_k, \; k=1,...,K,
\end{equation}
where $\hat{\h}_k\in\C^{1\times N}$ is the erroneous channel and $\eee_k\in\C^{1\times N}$ represents the additive CSI error, while only $\hat{\h}_k$ is assumed to be known at the transmitter. The actual channel $\h_k$, the estimate channel $\hat{\h}_k$, and the CSI error vector $\eee_k$ are assumed to be mutually uncorrelated for all $k=1,...,K$. In order to characterize the channel error vectors \(\{\eee_k\}^{K}_{^{}k=1^{}}\), we consider two different models as follows.

\subsection{Spherical Uncertainty Region}

The spherical uncertainty model assumes the actual channel $\h_k$ to always lie inside a sphere (in general, ellipsoid) centered at the erroneous channel $\hat{\h}_k$, with some known (deterministic) radius $\varepsilon_k$. In a formal way, it is assumed that $\h_k$ belongs to a spherical uncertainty set defined as

\begin{equation}\label{eq:unc}
\mathcal{H}_k\triangleq\left\{\h_k : \|\h_k-\hat{\h}_k\|_2\leq \varepsilon_k \right\},
\end{equation}
from which the $k$th actual channel is equally described by
\begin{equation}\label{eq:delc}
\h_k = \hat{\h}_k + \eee_k, \; \|\eee_k\|_2\leq \varepsilon_k.
\end{equation}
It is therefore clear that the uncertain component of the CSI in the spherical model \eqref{eq:delc} is a vector with a bounded norm. This model is particularly suitable for wireless systems with finite-rate feedback in which the CSI is acquired and quantized at the receiver and fed back to the transmitter \cite{feedback_mimo,maximin}. Notice that, in this model, usually no assumption is made on the distribution of $\eee_k$.
\subsection{Stochastic Uncertainty Region}

It is commonly assumed, in wireless scenarios with imperfect channel estimation, that the transmitter is only provided with an estimate channel $\hat{\h}_k$, while the vector $\eee_k$ captures the Gaussian estimation error. In this case, the $k$th actual channel is modeled as
\begin{equation}\label{eq:delc2}
\h_k = \hat{\h}_k + \eee_k, \; \eee_k\sim\mathcal{CN}(\OOO, \xi_k^2\,\I),
\end{equation}
where the error variance $\xi_k^2$ is known to the transmitter and generally depends on the quality of the estimate channel and the imperfections in the estimation process. The stochastic error model specifically corresponds to time-division duplex systems, where the transmitter exploits the estimated uplink channel for the downlink precoding \cite{chance_holger}. It is worth noting that the uncertainty model \eqref{eq:delc2} may also appear in a different scenario with statistical CSI in which the channel statistics are assumed to be (partially) known at the transmitter, in a way that either the channel's mean or covariance (or both) is (are) available; see, for example, \cite{palomar_thesis, unc_sta, csierror_goldsmith}. In such case, one may model the statistical CSI as $\h_k\sim\mathcal{CN}(\hat{\h}_k,\xi_k^2\,\I)$, which leads ultimately to similar results.

From now on, it is more convenient to use equivalent real-valued notations instead of the complex-valued ones, i.e., $$\tilde{\uuu}=\begin{bmatrix}\REAL(\uuu) \\ \IMAG(\uuu)\end{bmatrix}\in\mathbb{R}^{2N\times1},\quad\s_k=\begin{bmatrix}\REAL(s_k) \\ \IMAG(s_k)\end{bmatrix}\in\mathbb{R}^2,k=1,...,K.$$
Furthermore, by defining the operator
\begin{equation*}\label{eq:trans}
\TT(\x) \triangleq
\begin{bmatrix}
\REAL(\x) & -\IMAG(\x)\\
\IMAG(\x) & \;\;\;\REAL(\x)
\end{bmatrix},
\end{equation*}
for any given complex vector $\x$, we denote $$\HHH_k=\TT(\h_k),\;\hat{\HHH}_k=\TT(\hat{\h}_k),\;\EEE_k=\TT(\eee_k),\;k=1,...,K,$$
all belonging to $\mathbb{R}^{2\times2N}$. From the new notations above, it is immediately apparent that
\begin{equation}\label{eq:Hr}
\HHH_k = \hat{\HHH}_k + \EEE_k, \; k=1,...,K.
\end{equation}
Notice also that $\|\EEE_k\|_F \!\triangleq\! \sqrt{\TR(\EEE_k\EEE_k^T)}\!=\! \sqrt{2}\|\eee_k\|_2\!\leq\!\sqrt{2}\varepsilon_k$, in the spherical model \eqref{eq:delc}, and $\EEE_k(j,:)\sim\mathcal{N}(\OOO,\,\frac{1}{2}\xi_k^2\,\I),k=1,...,K,j=1,2$, in the stochastic model \eqref{eq:delc2}, where $\EEE_k(j,:)$ refers to the $j$th row of $\EEE_k$. In the rest of this paper, we unify the norm notations such that $\|\cdot\|$ denotes either the Frobenius norm of a matrix or the Euclidean norm of a vector. In addition, for each user $k=1,...,K$, by the received signal we mean the noise-free received signal, i.e., $\HHH_k\tilde{\uuu}$.



\section{Robust CI Formulation with Imperfect CSI}\label{sec:slp}

In the symbol-level precoding optimization, a crucial design constraint is to accommodate the received signal of each user $k$ into a pre-specified region, called constructive interference region (CIR), which corresponds to the intended symbol $s_k$. The CIRs, which are modulation-specific regions, have been defined in several ways in the literature; see, e.g., \cite{slp_chr,slp_con,slp_gen}. As mentioned earlier, we focus on the so-called distance-preserving CIRs (DPCIR) \cite{slp_gen}, which are defined in a generic form that is applicable to any given (two-dimensional) modulation scheme.

In a non-robust design, one may only rely on the estimate channels
$\{\hat{\HHH}_k\}_{k=1}^K$ in order to optimize the transmit signal $\tilde{\uuu}$. Let us first assume that the downlink channels are perfectly known to the transmitter, i.e., $\hat{\HHH}_k=\HHH_k, k=1,...,K$.
It has been shown in \cite{slp_tsp} that the distance-preserving CI constraints can be introduced in the precoding design problem in the form of vector inequalities
\begin{equation}\label{eq:cic}
\A_k \hat{\HHH}_k \tilde{\uuu} \geq \mu_k \A_k \s_k, \; k=1,...,K,
\end{equation}
where $\A_k\in\mathbb{R}^{2\times2}$ describes the distance-preserving region associated with $\s_k$ (notice that each symbol $\s_k$ corresponds to a constellation point), and $\mu_k$ is an amplitude scalar determined by the type of the design problem. As a specific example that corresponds to our design criterion, one may consider $\mu_k=\sigma_k \sqrt{\gamma_k}$ in the SINR-constrained power minimization problem, with $\gamma_k$ denoting the given SINR requirement of the $k$th user. Notice that ``SINR'' equally refers to ``SNR'' in the context of symbol-level precoding; see \cite{slp_tsp}. It is also worth mentioning that the matrix $\A_k$ contains the normal vectors of the two distance-preserving boundaries associated with symbol $\s_k$. More details on how to describe the DPCIRs as in \eqref{eq:cic} can be found in \cite{slp_tsp,slp_cf}.

With imperfect CSI, however, the regions described by \eqref{eq:cic} are distorted versions of the accurate CI regions. As a result, the received signals $\{\HHH_k\tilde{\uuu}\}_{k=1}^K$ are no longer guaranteed to lie in the desired CI regions, causing performance degradation, e.g., a higher symbol error probability. Therefore, in order for any robust design of symbol-level precoding, one first needs to properly reformulate the CI constraints in accordance with each uncertainty model.



The accurate CI constraint to be met for any user $k$ is 
\begin{equation}\label{eq:cicimpr}\nonumber
\A_k \HHH_k \tilde{\uuu} \geq \sigma_k\sqrt{\gamma_k} \A_k \s_k, \; k=1,...,K,
\end{equation}
By substituting \eqref{eq:Hr} for $\HHH_k$, we have
\begin{equation}\label{eq:cicimp2r}
\A_k \hat{\HHH}_k \tilde{\uuu} \geq \sigma_k\sqrt{\gamma_k} \A_k \s_k - \A_k \EEE_k \tilde{\uuu} , \; k=1,...,K.
\end{equation}
A robust CI constraint must aim to satisfy \eqref{eq:cicimp2r} for any possible realization of the CSI error $\EEE_k$ taken from the uncertainty set.
In the sequel, we separately consider each uncertainty model and derive robust formulation(s) for the CI constraints. For the brevity of notation, we hereafter denote by
\begin{equation}\label{eq:vec3}
\www_k(\tilde{\uuu}) \triangleq \sigma_k\sqrt{\gamma_k} \A_k \s_k - \A_k \hat{\HHH}_k \tilde{\uuu},
\end{equation}
the certain part of the CI inequality  \eqref{eq:cicimp2r} which is affine in $\tilde{\uuu}$, where $\www_k(\tilde{\uuu})=[w_{k,1},w_{k,2}]^T$.

\subsection{Worst-case Robust Formulation}

The spherical (norm-bounded) uncertainty region $\mathcal{H}_k$ can be interpreted as having all the possible error vectors inside a $2N$-dimensional sphere with radius $\sqrt{2}\varepsilon_k$. In this case, the robust formulation of \eqref{eq:cicimp2r} for the $k$th user can be written as
\begin{equation}\label{eq:pmall}
\begin{aligned}
\A_k \EEE_k \tilde{\uuu} \geq \www_k(\tilde{\uuu}), \; \forall \EEE_k : \|\EEE_k\| \leq \sqrt{2} \, \varepsilon_k,
\end{aligned}
\end{equation}
which implies that \eqref{eq:cicimp2r} must be satisfied for all $\EEE_k$ belonging to the CSI uncertainty set. Even though the feasibility region of \eqref{eq:pmall} is convex, this semi-infinite constraint consists of an infinite number of linear inequalities to be satisfied which is computationally intractable.
In order to achieve robustness over a bounded uncertainty set as in \eqref{eq:pmall}, a common approach is to consider the design constraint in its worst case. Accordingly, letting $\A_k = [\aaa_{k,1},\aaa_{k,2}]^T$, the worst-case formulation of \eqref{eq:pmall} can be written as
\begin{equation}\label{eq:pmwc}
\begin{aligned}
\begin{bmatrix} \INF \{\aaa_{k,1}^T \EEE_k \tilde{\uuu} : \|\EEE_k\| \leq \sqrt{2} \, \varepsilon_k\} \\ \INF \{\aaa_{k,2}^T \EEE_k \tilde{\uuu} : \|\EEE_k\| \leq \sqrt{2} \, \varepsilon_k\} \end{bmatrix} \geq \www_k(\tilde{\uuu}).
\end{aligned}
\end{equation}
In our model, the worst-case uncertainty is realized through the maximal CSI error norm, i.e., the radius of the CSI error sphere. From the definition of the spherical uncertainty set in \eqref{eq:unc}, it can be easily shown that the entries of  $\A_k \EEE_k \tilde{\uuu}$ are bounded too. We also remark that
\begin{equation}\label{eq:vec}
\A_k \EEE_k \tilde{\uuu} = (\tilde{\uuu}^T \otimes \A_k) \, \VEC(\EEE_k),
\end{equation}
which can be simply verified using the well-known property $\VEC(\XXX\YYY\WWW)=(\WWW^T \otimes \XXX)\,\VEC(\YYY)$, for any given matrices $\XXX,\YYY,\WWW$ with appropriate dimensions, and also the fact that $\A_k \EEE_k \tilde{\uuu}=\VEC(\A_k \EEE_k \tilde{\uuu})$. It then follows that
\begin{equation}\label{eq:vk}
\A_k \EEE_k \tilde{\uuu} = \begin{bmatrix}
(\tilde{\uuu}^T \otimes \aaa_{k,1}^T) \, \VEC(\EEE_k) \\
(\tilde{\uuu}^T \otimes \aaa_{k,2}^T) \, \VEC(\EEE_k)
\end{bmatrix}.
\end{equation}
Now, let us focus on the rows of the right-hand side vector in \eqref{eq:vk}. By the Cauchy-Schwarz inequality, we have
\begin{equation}\label{eq:cs}
\begin{aligned}
(\tilde{\uuu}^T \otimes \aaa_{k,j}^T) \VEC(\EEE_k) \geq -\|\tilde{\uuu}^T \otimes \aaa_{k,j}^T\|\,\|\VEC(\EEE_k)\|, \, j=1,2.
\end{aligned}
\end{equation}
Using the uncertainty radius $\|\VEC(\EEE_k)\| = \|\EEE_k\| \leq \sqrt{2}\,\varepsilon_k$, an immediate consequence of \eqref{eq:cs} is that $(\tilde{\uuu}^T \otimes \aaa_{k,j}^T) \VEC(\EEE_k)$ is bounded from below by $-\sqrt{2} \, \varepsilon_k \, \|\tilde{\uuu}^T \otimes \aaa_{k,j}^T\|$ for $j=1,2$. However, by exploiting the structure of $\VEC(\EEE_k)$, it is possible to further obtain a tighter bound which is given by
\begin{equation}\label{eq:supp}
\begin{aligned}
\INF \, \Big\{(\tilde{\uuu}^T \otimes \aaa_{k,j}^T) \, \VEC(\EEE_k) :\, \|\EEE_k\| \leq \sqrt{2}\varepsilon_k \Big\} = -\varepsilon_k \, \|\tilde{\uuu}^T \otimes \aaa_{k,j}^T\|
= -\varepsilon_k \, \|\tilde{\uuu}\| \, \|\aaa_{k,j}\|, \; j=1,2,
\end{aligned}
\end{equation}
where the last equality of \eqref{eq:supp} is derived considering the fact that $\|\x \otimes \yyy\|=\|\x\| \, \|\yyy\|$, for any two vectors $\x$ and $\yyy$. Finally, substituting \eqref{eq:supp} for the infimum in \eqref{eq:pmwc}, the worst-case CI constraint for the $k$th user is obtained by
\begin{equation}\label{eq:pmwc2}
\begin{aligned}
-\varepsilon_k \, \|\tilde{\uuu}\| \begin{bmatrix} \|\aaa_{k,1}\| \\ \|\aaa_{k,2}\| \end{bmatrix} \geq \www_k(\tilde{\uuu}),
\end{aligned}
\end{equation}
The CI constraint \eqref{eq:pmwc2} can be equivalently expressed by two second-order cone (SOC) constraints, given in a compact form by
\begin{equation}\label{eq:pmwc3}
\begin{aligned}
\mathrm{W}: \quad \|\tilde{\uuu}\| \, \pmb{1} \leq \frac{-1}{\varepsilon_k} (\A_k \A_k^T \circ \I)^{-1/2} \www_k(\tilde{\uuu}).
\end{aligned}
\end{equation}
In fact, the worst-case constraint $\mathrm{W}$ guarantees that the CI requirement for the $k$th user will be met in the presence of any unknown, but norm-bounded CSI error. The robust formulation \eqref{eq:pmwc3} is convex and thus can efficiently be handled via off-the-shelf convex optimization algorithms\cite{convex_boyd}.
It is worth mentioning that a similar worst-case robust approach has also been studied in \cite{slp_chr} for symbol-level downlink precoding in which the CI regions coincide with the DPCIRs in the special case of PSK signaling, but characterization of the CI constraints are not identical. Nevertheless, the final robust formulations, despite being different in presentation, are based on the same idea and are basically equivalent. 
\subsection{Stochastic Robust Formulation}

Assuming statistically-known CSI errors, the CI constraint in \eqref{eq:cicimp2r} turns into an uncertain inequality with the uncertainty arising from the stochastic CSI error $\EEE_k$. Although the feasible set of this uncertain inequality is always convex, the major difficulty is to efficiently check whether this convex constraint is satisfied at a given point, which is highly computationally demanding. In such case, the (deterministic) constraint in \eqref{eq:cicimp2r} can be reformulated as a probabilistic constraint (commonly known as chance constraint). The chance constraint then implies that the $k$th user will experience the event of CI failure only with a constrained small probability, i.e.,
\begin{equation}\label{eq:pmr}
\begin{aligned}
\PR\left\{\A_k \hat{\HHH}_k \tilde{\uuu} \ngeq \sigma_k\sqrt{\gamma_k} \A_k \s_k - \A_k \EEE_k \tilde{\uuu}\right\} < \upsilon,
\end{aligned}
\end{equation}
which can be equally expressed by
\begin{equation}\label{eq:pmr2}
\begin{aligned}
\PR\left\{\A_k \hat{\HHH}_k \tilde{\uuu} \geq \sigma_k\sqrt{\gamma_k} \A_k \s_k - \A_k \EEE_k \tilde{\uuu}\right\} \geq 1-\upsilon,
\end{aligned}
\end{equation}
where $\upsilon\in(0,1/2]$ denotes the violation probability threshold which is a system design parameter controlling the desired level of conservatism. Remark that the SINR requirement $\gamma_k$ translates to an achievable rate target of $R_k=\log_2(1+\gamma_k)$, under ergodic conditions on the channel \cite{tse}; therefore, the constraint \eqref{eq:pmr2} can also be read as a rate-outage probability constraint, ensuring that the transmission rate $R_k$ is achievable for the $k$th user with probability (at least) $1-\upsilon$. For the sake of notation, we denote by
\begin{equation}\label{eq:vec2}
\qqq_k \triangleq \A_k \EEE_k \tilde{\uuu} = (\tilde{\uuu}^T \otimes \A_k) \, \VEC(\EEE_k),
\end{equation}
the stochastic uncertain component of the CI constraint, where $\qqq_k=[q_{k,1},q_{k,2}]^T$. The chance constraint \eqref{eq:pmr2} can then be written, in a simpler form, as
\begin{equation}\label{eq:prob}
\PR\left\{\qqq_k \geq \www_k(\tilde{\uuu}) \right\} \geq 1-\upsilon, \; k=1,...,K.
\end{equation}
The constraints in \eqref{eq:prob} belong to chance-constrained vector inequalities, which are generally known to be computationally intractable \cite{safe}, as we will also see later. In what follows, the goal is to derive equivalent deterministic expressions for \eqref{eq:prob}.
For this purpose, we first need to study the statistical properties of the uncertain vector $\qqq_k$.

We begin with the Gaussian error vector $\VEC(\EEE_k)$, which can be identified by its mean and covariance matrix given by $\EXP\{\VEC(\EEE_k)\} = \; \OOO$
and
\begin{equation}\label{Edel}
\EXP\left\{\VEC(\EEE_k)\VEC(\EEE_k)^T\right\} = \frac{1}{2} \, \xi_k^2\begin{bmatrix} \I_{2N} & \J \\
\J^T & \I_{2N}
\end{bmatrix},
\end{equation}
respectively, where $$\J=\I_N \otimes \J_2, \; \J_2 \triangleq \begin{bmatrix} 0 & 1 \\ -1 & 0
\end{bmatrix}.$$
From \eqref{eq:vec2}, it is straightforward to show that $\qqq_k$ is a (possibly correlated) Gaussian random vector with mean
\begin{equation}\label{eq:vkm}
\begin{aligned}
\EXP\{\qqq_k\} = \left(\tilde{\uuu}^T \otimes \A_k\right) \, \EXP\left\{\VEC(\EEE_k)\right\} = \OOO,
\end{aligned}
\end{equation}
and covariance
\begin{equation}\label{eq:vkv}
\begin{aligned}
\CCC_k = \EXP\{\qqq_k \qqq_k^T\}
 \overset{(\mathrm{a})}{=} (\tilde{\uuu}^T \otimes \A_k) \; \EXP\Big\{\VEC(\EEE_k)\VEC(\EEE_k)^T\Big\} (\tilde{\uuu} \otimes \A_k^T)
 \overset{(\mathrm{b})}{=} \frac{1}{2} \, \xi_k^2 \, \left(\tilde{\uuu}^T \tilde{\uuu} \otimes \A_k \A_k^T\right) = \frac{1}{2} \, \xi_k^2 \, \|\tilde{\uuu}\|^2 \, \A_k \A_k^T,
\end{aligned}
\end{equation}
where the equality (a) is verifiable by using the property $(\XXX \otimes \YYY)^T=(\XXX^T \otimes \YYY^T)$, for any given matrices $\XXX,\YYY,\WWW,\ZZZ$, and the equality (b) has been verified in Appendix \ref{app:b}.
Using the first two moments of $\qqq_k$, the probability in \eqref{eq:prob} can be precisely evaluated as the integral of the joint Gaussian probability distribution of $q_{k,1}$ and $q_{k,2}$, i.e.,
\begin{equation}\label{eq:int}
\begin{aligned}
\PR \{ \qqq_k &\geq \www_k(\tilde{\uuu})\} = \PR\left\{q_{k,1} \geq w_{k,1},q_{k,2} \geq w_{k,2}\right\}
= \int\limits_{w_{k,2}}^{\infty}\int\limits_{w_{k,1}}^{\infty}\frac{1}{2\pi\sqrt{|\CCC_k|}}  \exp\!\left\{\!-\frac{1}{2}\qqq_k^T\CCC_k^{-1}\qqq_k\right\}\! \mathrm{d}q_{k,1} \mathrm{d}q_{k,2}.
\end{aligned}
\end{equation}
However, no explicit closed-form expression is known for the integral in \eqref{eq:int}. It becomes even more challenging to imply the constraint \eqref{eq:int} in the precoding optimization problem. In order to resolve the difficulty of finding a tractable (convex) expression for \eqref{eq:int}, a straightforward approach is to eliminate the (possible) correlation between the entries of $\qqq_k$ through applying a whitening transform. In this regard, the optimal whitening matrix (in the sense of minimum mean-square error) is shown in \cite{white} to be
\begin{equation}\label{eq:wmat}
\begin{aligned}
\CCC_k^{-1/2} = \frac{\sqrt{2}}{\xi_k \, \|\tilde{\uuu}\|} \, (\A_k \A_k^T)^{-1/2},
\end{aligned}
\end{equation}
where $(\cdot)^{-1/2}$ denotes the inverse square root. It is worthwhile to mention that in \cite{slp_tsp}, the $2\times2$ matrix $\A_k \A_k^T$ is proven to be always non-singular. Hence, $\CCC_k$ is positive definite and has a unique (invertible) square root. As a result, the probability \eqref{eq:int} can be equally expressed by
\begin{equation}\label{eq:dec}
\begin{aligned}
\PR \left\{\qqq_k \geq \www_k(\tilde{\uuu})\right\} = \PR \left\{\CCC_k^{1/2}\CCC_k^{-1/2}\qqq_k \geq \www_k(\tilde{\uuu})\right\}
&= \PR \left\{\bar{\qqq}_k \geq \CCC_k^{-1/2}\www_k(\tilde{\uuu})\right\} = \PR \left\{\bar{\qqq}_k \geq \bar{\www}_k(\tilde{\uuu})\right\}\!,
\end{aligned}
\end{equation}
where $\bar{\qqq}_k\triangleq\CCC_k^{-1/2}\qqq_k$ and $\bar{\www}_k(\tilde{\uuu})\triangleq\CCC_k^{-1/2}\www_k(\tilde{\uuu})$.
It can be easily verified that $\bar{\qqq}_k$ is an uncorrelated zero-mean Gaussian random vector with unit diagonal covariance matrix, i.e.,
\begin{equation}\label{eq:white}
\begin{aligned}
\bar{\CCC}_k \triangleq \EXP\left\{\bar{\qqq}_k \bar{\qqq}_k^T\right\}
= \EXP\left\{\CCC_k^{-1/2}\qqq_k \qqq_k^T \CCC_k^{-1/2}\right\}
= \CCC_k^{-1/2}\EXP\left\{\qqq_k \qqq_k^T\right\}\CCC_k^{-1/2} = \CCC_k^{-1/2} \CCC_k \CCC_k^{-1/2} = \I.
\end{aligned}
\end{equation}
Consequently, the chance constraint \eqref{eq:prob} boils down to
\begin{equation}\label{eq:probunc}
\PR\left\{\bar{\qqq}_k \geq \bar{\www}_k(\tilde{\uuu}) \right\} \geq 1-\upsilon,
\end{equation}
with $\bar{\qqq}_k \sim \mathcal{N}(\OOO,\I)$. This probability may appear to be easily handled as it can be expressed by the product of two (complementary) error functions. In the context of convex optimization, however, we essentially need to reach a convex representation for \eqref{eq:probunc}. This could be in general an intricate task since the joint probability in \eqref{eq:probunc} does not admit a tractable convex expression. An alternative approach to tackle this intractability is to replace \eqref{eq:probunc} with its {\emph{safe tractable approximation}}, resulting in an efficiently computable convex constraint. Such an approximation lies within the literature of robust optimization techniques \cite{safe,safe2}. The term {\emph{safe}} is used here in the sense that the feasible points of the safe approximation must be necessarily feasible also for \eqref{eq:probunc}. Therefore, in what follows the goal is to propose computationally tractable (but possibly not equivalent) convex approximations implying the CI chance constraint \eqref{eq:probunc}.

\noindent{\bf{Remark 1.}} Using the fact that $\bar{\qqq}_k$ has a symmetric distribution, it is trivial to show that the chance constraint \eqref{eq:probunc} is feasible for every $\upsilon\in(0,1/2]$ if and only if we have $\EXP\{\bar{\qqq}_k\} \geq \bar{\www}_k(\tilde{\uuu})$. Consequently, under the assumption $\upsilon\in(0,1/2]$, a necessary and sufficient condition for \eqref{eq:probunc} to have a nonempty feasible set is $\bar{\www}_k(\tilde{\uuu}) \leq \OOO$.


\subsubsection{Safe Approximation I}
One may simply exploit the fact that the two random entries of $\bar{\qqq}_k$ are uncorrelated, hence independent. Consequently, denoting $\bar{\qqq}_k=[\bar{q}_{k,1},\bar{q}_{k,2}]^T$ and $\bar{\www}_k(\tilde{\uuu})=[\bar{w}_{k,1},\bar{w}_{k,2}]^T$, by using the Gaussian cumulative distribution function, the joint probability in \eqref{eq:probunc} can be separated as
\begin{equation}\label{eq:erf}
\begin{aligned}
\PR\left\{\bar{\qqq}_k \geq \bar{\www}_k(\tilde{\uuu}) \right\} = \PR \left\{\bar{q}_{k,1} \geq \bar{w}_{k,1}\right\} \; \PR \left\{\bar{q}_{k,2} \geq \bar{w}_{k,2}\right\}
= \frac{1}{2} \mathrm{erfc}\left(\frac{\bar{w}_{k,1}}{\sqrt{2}}\right) \times \frac{1}{2} \mathrm{erfc}\left(\frac{\bar{w}_{k,2}}{\sqrt{2}}\right),
\end{aligned}
\end{equation}
where $\mathrm{erfc}(\cdot)$ is the complementary error function defined by $\mathrm{erfc}(z)\triangleq\frac{2}{\sqrt{\pi}}\int_z^\infty e^{-t^2} dt$. Due to the decreasing monotonicity of the complementary error function, the desired probability is always bounded from below by
\begin{equation}\label{eq:erflb}
\begin{aligned}
\PR\left\{\bar{\qqq}_k \geq \bar{\www}_k(\tilde{\uuu}) \right\} &\geq \frac{1}{4} \, \mathrm{erfc}^2\left(\frac{\max\{\bar{w}_{k,1},\bar{w}_{k,2}\}}{\sqrt{2}}\right).
\end{aligned}
\end{equation}
Using \eqref{eq:erflb}, in order to imply the chance constraint \eqref{eq:probunc}, it is sufficient to consider the deterministic constraint
\begin{equation}\label{eq:prob2}
\begin{aligned}
\frac{1}{4} \, \mathrm{erfc}^2\left(\frac{\max\{\bar{w}_{k,1},\bar{w}_{k,2}\}}{\sqrt{2}}\right) \geq 1-\upsilon,
\end{aligned}
\end{equation}
which can be written as
\begin{equation}\label{eq:lmi}
\quad -\max\left[\bar{\www}_k(\tilde{\uuu})\right] \leq \rho(\upsilon),
\end{equation}
where $\rho(\upsilon) \triangleq -\sqrt{2} \,\mathrm{erfc}^{-1}\left(2 \sqrt{1 - \upsilon}\right)$ with $\mathrm{erfc}^{-1}\!(\cdot)$ denoting the inverse complementary error function, and $\max[\cdot]$ is the entrywise maximum. By replacing $\bar{\www}_k(\tilde{\uuu})$, the conservative robust approximation \eqref{eq:lmi} can be rewritten as an SOC constraint
\begin{equation}\label{eq:lmi2}
\mathrm{A}1 : \quad \|\tilde{\uuu}\| \leq \frac{-\sqrt{2}}{\rho(\upsilon) \, \xi_k} \max\left[(\A_k \A_k^T)^{-1/2} \www_k(\tilde{\uuu})\right],
\end{equation}
It should be remarked that, in general, the feasible region of $\mathrm{A}1$ is a convex subset of that of \eqref{eq:probunc}. Therefore, the convex approximation $\mathrm{A}1$ may not exactly imply the desired chance constraint \eqref{eq:probunc}, but any feasible solution to \eqref{eq:lmi2} is guaranteed to be feasible also for \eqref{eq:probunc}.


\subsubsection{Safe Approximation II} Our subsequent derivation of a second safe tractable approximation for \eqref{eq:probunc} is essentially based on the well-known Schur complement lemma and the following theorem \cite[Th. 4.1]{safe}.
\begin{lemma}\label{lem:1} (Schur complement)
        Let $\WWW$ be a symmetric matrix given by
        \begin{equation}\label{eq:schur}
        \WWW = \begin{bmatrix} \XXX & \YYY \\ \YYY^T & \ZZZ \end{bmatrix}.
        \end{equation}
        Then, $\WWW \succeq 0$ if and only if $\XXX \succeq 0$ and $\Del_\XXX \succeq 0$, where $\Del_\XXX = \ZZZ - \YYY^T \XXX^{-1} \YYY$ is the Schur complement of $\XXX$ in $\WWW$.
\end{lemma}

\begin{theorem}\label{thm:1}
        Let $\DDD_0,\DDD_1,...,\DDD_L$ be diagonal $n \times n$ matrices with $\DDD_0 \succeq 0$, and $\zeta_1,...,\zeta_L$ be mutually independent random variables where $\zeta_l \sim \mathcal{N}(0,1),\forall l\in\{1,...,L\}$. Then, the semidefinite constraint
        $$\mathrm{Arw}\left(\DDD_0,\DDD_1,...,\DDD_L\right)\succeq0,$$
        implies, for every $\upsilon\in(0,1/2]$, that
        $$\PR\left\{-\psi(\upsilon) \DDD_0 \preceq \sum_{l=1}^L \zeta_l \DDD_l \preceq \psi(\upsilon) \DDD_0\right\}\geq 1 - \upsilon,$$
        with $\psi(\upsilon) = \mathrm{erfc}^{-1}\left(\frac{\upsilon}{2n}\right)$, where
        $$\mathrm{Arw}\left(\DDD_0,\DDD_1,...,\DDD_L\right)\triangleq\begin{bmatrix} \DDD_0 & \DDD_1 & \DDD_2 & \cdots & \DDD_L\\
        \DDD_1 & \DDD_0 & \OOO & \cdots & \OOO \\
        \DDD_2 & \OOO & \DDD_0 & \cdots & \OOO \\
        \vdots & \vdots & \vdots & \ddots & \vdots \\
        \DDD_L & \OOO & \OOO & \cdots & \DDD_0
        \end{bmatrix}.$$
\end{theorem}

We recall that our goal here is to find a tractable sufficient (convex) condition for the CI inequality in \eqref{eq:probunc} to be satisfied with probability at least $1-\upsilon$. The inequality of interest, i.e., $\bar{\qqq}_k \geq \bar{\www}_k(\tilde{\uuu})$, can be equivalently expressed by a linear matrix inequality (LMI) as
\begin{equation}\label{eq:probunc2}
\psi(\upsilon) \DDD_{0,k} + \bar{q}_{k,1} \DDD_1 + \bar{q}_{k,2} \DDD_2 \succeq 0,
\end{equation}
where
$$\DDD_{0,k} \!\triangleq\! \frac{1}{\psi(\upsilon)} \begin{bmatrix}
-\bar{w}_{k,1} & 0 \\ 0 & -\bar{w}_{k,2}
\end{bmatrix},\DDD_1 \!\triangleq\! \begin{bmatrix}
1 & 0 \\ 0 & 0
\end{bmatrix},\DDD_2 \!\triangleq\! \begin{bmatrix}
0 & 0 \\ 0 & 1
\end{bmatrix},$$
Since $\bar{q}_{k,1}$ and $\bar{q}_{k,2}$ are both symmetric in distribution and the violation probability $\upsilon$ is (typically) small, a sufficient condition for
\begin{equation}\label{eq:probunc22}
\PR\left\{\psi(\upsilon) \DDD_{0,k} + \bar{q}_{k,1} \DDD_1 + \bar{q}_{k,2} \DDD_2 \succeq 0\right\}\geq 1-\upsilon,
\end{equation}
is also sufficient for
\begin{equation}\label{eq:probunc3}
\PR\left\{-\psi(\upsilon) \DDD_{0,k} \preceq \bar{q}_{k,1} \DDD_1 + \bar{q}_{k,2} \DDD_2 \preceq \psi(\upsilon) \DDD_{0,k}\right\}\geq 1-\upsilon.
\end{equation}
By a direct application of Theorem \ref{thm:1} with $n=2$ and $L=2$, it follows that the chance constraint \eqref{eq:probunc3}
is met if
\begin{equation}\label{eq:safe2}
\mathrm{Arw}(\DDD_{0,k},\DDD_1,\DDD_2)\succeq 0,
\end{equation}
holds true with $\psi(\upsilon) = \mathrm{erfc}^{-1}\left(\frac{\upsilon}{4}\right)$.
Notice that a necessary condition for Theorem \ref{thm:1} to be valid is $\DDD_{0,k} \succeq 0$. The matrix $\mathrm{Arw}(\DDD_{0,k},\DDD_1,\DDD_2)$ is symmetric, and further, can be partitioned as required in \eqref{eq:schur}. As a result, using Lemma \ref{lem:1} with $\XXX=\DDD_{0,k}$ and $\WWW=\mathrm{Arw}(\DDD_{0,k},\DDD_1,\DDD_2)$, it can be immediately verified that the following implication holds:
\begin{equation}\label{eq:safe4}
\mathrm{Arw}(\DDD_{0,k},\DDD_1,\DDD_2)\succeq 0 \implies \DDD_{0,k} \succeq 0.
\end{equation}
Therefore, the safe convex constraint \eqref{eq:safe2} sufficiently implies our desired chance constraint in \eqref{eq:probunc3}. Finally, by replacing $\DDD_{0,k}$, $\DDD_1$ and $\DDD_2$ in \eqref{eq:safe2}, the safe convex approximation is obtained as the semidefinite constraint
\begin{equation}\label{eq:safe3}
\setlength\arraycolsep{1pt}
\begin{bmatrix}
-\frac{\bar{w}_{k,1}}{\psi(\upsilon)} & 0 & 1 & 0 & 0 & 0 \\
0 & -\frac{\bar{w}_{k,2}}{\psi(\upsilon)} & 0 & 0 & 0 & 1 \\
1 & 0 & -\frac{\bar{w}_{k,1}}{\psi(\upsilon)} & 0 & 0 & 0 \\
0 & 0 & 0 & -\frac{\bar{w}_{k,2}}{\psi(\upsilon)} & 0 & 0 \\
0 & 0 & 0 & 0 & -\frac{\bar{w}_{k,1}}{\psi(\upsilon)} & 0 \\
0 & 1 & 0 & 0 & 0 & -\frac{\bar{w}_{k,2}}{\psi(\upsilon)}
\end{bmatrix} \succeq 0.
\end{equation}
It is routine to check that the LMI in \eqref{eq:safe3} is not convex in the given form with respect to $\tilde{\uuu}$. Nevertheless, it has been shown in Appendix B that, using the implication provided in Remark 1, it is possible to recast the semidefinite constraint \eqref{eq:safe3} as an equivalent SOC constraint given by
\begin{equation}\label{eq:safe6}
\mathrm{A}2: \quad \|\tilde{\uuu}\| \, \pmb{1} \leq \frac{-\sqrt{2}}{\psi(\upsilon) \, \xi_k}(\A_k \A_k^T)^{-1/2} \www_k(\tilde{\uuu}),
\end{equation}
which is indeed convex in $\tilde{\uuu}$, and can efficiently be handled by standard convex optimization solvers \cite{convex_boyd}.
 
In order to gain some insight into the proposed safe convex approximation $\mathrm{A}2$, and further for comparison purposes, we also formulate a benchmark approximation based on the so-called sphere bounding method. The idea (in some sense) is borrowed from the worst-case robust design approach. More specifically, the goal is basically to find a bounded uncertainty set to which the stochastically-uncertain component in \eqref{eq:probunc} belongs with a certain probability; subsequently, the worst-case approach can be applied. The following lemma from \cite{outage_conic} helps us to proceed with the formulation.
\begin{lemma}\label{lem:2}
        Let $\mathcal{S}\subset\mathbb{R}^n$ be an arbitrary set with the property $f(\x) \geq \OOO,\forall\x\in\mathcal{S}$, where $f(\cdot)$ is in general a vector-valued function. Then, for a given $\yyy \in \mathbb{R}^n$, the restriction $$\PR\left\{f(\yyy) \geq \OOO\right\} \geq 1-\upsilon,$$ is implied sufficiently by satisfying $\PR\left\{\yyy \in \mathcal{S}\right\} \geq 1-\upsilon.$
\end{lemma}

In order to imply the chance constraint \eqref{eq:probunc}, one may use the implication provided by Lemma \ref{lem:2} to obtain a (preferably) tight convex restriction, as long as the resulting constraint is efficiently computable. This requires to properly choose the set $\mathcal{S} \subseteq \mathbb{R}^2$ in such a way that the condition
\begin{equation}\label{eq:cond1}
f(\bar{\qqq}_k)\geq \OOO, \; f(\bar{\qqq}_k)\triangleq\bar{\qqq}_k - \bar{\www}_k(\tilde{\uuu}),
\end{equation}
is met for all $\bar{\qqq}_k \in \mathcal{S}$, while satisfying $\PR\left\{\bar{\qqq}_k \in \mathcal{S} \right\} \geq 1-\upsilon$. We recall that $\bar{\qqq}_k \sim \mathcal{N}(\OOO,\I)$, and that $\bar{\qqq}_k$ has a symmetric distribution. Thus, the condition \eqref{eq:cond1} can be equally expressed as
\begin{equation}\label{eq:cond2}
f(\bar{\qqq}_k) \leq \OOO, \; f(\bar{\qqq}_k)\triangleq\bar{\qqq}_k + \bar{\www}_k(\tilde{\uuu}).
\end{equation}
A common (convex) choice for the set $\mathcal{S}$ to reach a computationally tractable formulation is the ball represented by
\begin{equation}\label{eq:ball}
\mathcal{S} \triangleq \left\{\x \in \mathbb{R}^2 : \|\x\| \leq \alpha(\upsilon)\right\},
\end{equation}
with a radius of $$\alpha(\upsilon) = \sqrt{\Phi^{-1}_2(1-\upsilon)}\,,$$ where $\Phi^{-1}_n(\cdot)$ is the inverse cumulative distribution function of the central Chi-square random variable with $n$ degrees of freedom. It is then straightforward to verify that
\begin{equation}\label{eq:ball2}
\PR\left\{\bar{\qqq}_k \in \mathcal{S} \right\} = 1-\upsilon,
\end{equation}
from which it can be presumed that $\bar{\qqq}_k$ is norm-bounded by $\alpha(\upsilon)$ with a probability of $1-\upsilon$. As a result,
\begin{equation}\label{eq:cond3}
\alpha(\upsilon) \pmb{1} + \bar{\www}_k(\tilde{\uuu}) \leq \OOO,
\end{equation}
implies that \eqref{eq:cond2} holds true for all $\bar{\qqq}_k \in \mathcal{S}$. Finally, the worst-case robust approximation \eqref{eq:cond3} can be expressed by an SOC constraint as
\begin{equation}\label{eq:cond4}
\mathrm{B}: \quad \|\tilde{\uuu}\| \, \pmb{1} \leq \frac{-\sqrt{2}}{\alpha(\upsilon) \, \xi_k} (\A_k \A_k^T)^{-1/2} \www_k(\tilde{\uuu}).
\end{equation}
In particular, the convex approximation $\mathcal{R}$ is able to control the radius $\alpha(\upsilon)$ according to the tolerable violation probability. It can immediately be inferred by comparing \eqref{eq:safe6} and \eqref{eq:cond4} that $\mathrm{A}2$ resembles the sphere bounding based approximation $\mathrm{B}$ in form. Based on this resemblance, the safe approximation method for $\upsilon\in(0,1/2]$ can be treated as defining the convex set $\mathcal{S}$ as a ball with a radius different from $\alpha(\upsilon)$, therefore with a different level of conservatism. In the next subsection, we compare the tightness of the proposed approximations with respect to the sphere bounding approach.


\subsection{Relative Tightness Comparison}

So far in this section, we have derived tractable convex formulations that, though not exact, sufficiently ensure the robust CI constraint of interest. This tractability led us to sacrifice tightness with respect to the originally intractable chance constraint \eqref{eq:probunc}. It is therefore desirable to investigate which formulation provides the tightest approximation among all the other ones.

\begin{table}[t]
        \caption{Summary of the proposed robust CI formulations.}
        \label{tab:c}
        \centering
        \renewcommand{\arraystretch}{1}
        \begin{tabular}{ll}
                \toprule
                Method & Robust CI constraint ($\forall k=1,...,K$)\\
                \midrule
                Worst-case & $\mathrm{W}\;: \enspace \|\tilde{\uuu}\| \pmb{1} \leq \frac{-1}{\varepsilon_k}(\A_k \A_k^T \circ \I)^{-1/2} \www_k(\tilde{\uuu})$\\
                & \qquad\enspace\, where $\www_k(\tilde{\uuu}) = \sigma_k\sqrt{\gamma_k} \A_k \s_k - \A_k \hat{\HHH}_k \tilde{\uuu}$\\
                \midrule
                Safe Approx. I & $\mathrm{A}1: \enspace \|\tilde{\uuu}\| \leq \frac{-\sqrt{2}}{\rho(\upsilon) \, \xi_k} \max\left[(\A_k \A_k^T)^{-1/2} \www_k(\tilde{\uuu})\right]$\\
                & \qquad\enspace\, with $\rho(\upsilon) = -\sqrt{2} \,\mathrm{erfc}^{-1}\!\left(2 \sqrt{1 - \upsilon}\right)$\\
                \midrule
                Safe Approx. II & $\mathrm{A}2: \enspace \|\tilde{\uuu}\|  \pmb{1} \leq \frac{-\sqrt{2}}{\psi(\upsilon) \, \xi_k} (\A_k \A_k^T)^{-1/2} \www_k(\tilde{\uuu})$\\
                & \qquad\enspace\, with $\psi(\upsilon) = \mathrm{erfc}^{-1}\left(\frac{\upsilon}{4}\right)$\\
                \midrule
                Sphere Bounding & $\mathrm{B}\;\;: \enspace \|\tilde{\uuu}\| \pmb{1} \leq \frac{-\sqrt{2}}{\alpha(\upsilon) \, \xi_k} (\A_k \A_k^T)^{-1/2} \www_k(\tilde{\uuu})$\\
                & \qquad\enspace\, with $\alpha(\upsilon) = \sqrt{\Phi^{-1}_2(1-\upsilon)}$\\
                \bottomrule
        \end{tabular}
\end{table}

Having rather similar conic representations for the three stochastic robust CI constraints, which are summarized in Table \ref{tab:c}, enables us to compare the relative tightness of the derived convex approximations. Here, we specifically define the relative tightness from the transmit power point of view according to which a convex approximation is a tighter one if it admits lower optimal transmit powers $\|\tilde{\uuu}\|^2$. We use the following two lemmas in the sequel. The proofs are straightforward.
\begin{lemma}\label{lem:3.5}
Let $\tilde{\uuu}^*$ be feasible to
        \begin{equation}\label{eq:tight0.5}
    \begin{aligned}
    \|\tilde{\uuu}\| \, \pmb{1} \leq \frac{-\sqrt{2}}{\beta \, \xi_k} (\A_k \A_k^T)^{-1/2} \www_k(\tilde{\uuu}),
    \end{aligned}
    \end{equation}
    with $\beta>0$, and satisfy $\bar{\www}_k(\tilde{\uuu}^*) \leq \OOO$ as a necessary condition. Then, it is implied that
        \begin{equation}\label{eq:tight1.5}
        \begin{aligned}
    \|\tilde{\uuu}^*\| \leq \frac{-\sqrt{2}}{\beta \, \xi_k} \max\left[(\A_k \A_k^T)^{-1/2} \www_k(\tilde{\uuu}^*)\right]
        \end{aligned}
        \end{equation}
        where $\max[\,\cdot\,]$ is the entrywise maximum of an input vector.
\end{lemma}
\begin{lemma}\label{lem:4}
        Consider the constraint
        \begin{equation}\label{eq:tight1}
        \|\tilde{\uuu}\| \leq \frac{-\sqrt{2}}{\beta \, \xi_k} \, \max\left[(\A_k \A_k^T)^{-1/2} \www_k(\tilde{\uuu})\right].
        \end{equation}
where $\beta>0$. Let $\tilde{\uuu}^*$ be feasible to \eqref{eq:tight1} with $\beta=\beta_1>0$, then for any $\beta_1 \geq \beta_2 > 0$, the following chain of inequalities holds:
\begin{equation}\label{eq:tight2}
\begin{aligned}
\|\tilde{\uuu}^*\| &\leq \frac{-\sqrt{2}}{\beta_1 \, \xi_k}\max\left[(\A_k \A_k^T)^{-1/2} \www_k(\tilde{\uuu}^*)\right]
\leq \frac{-\sqrt{2}}{\beta_2 \, \xi_k}\max\left[(\A_k \A_k^T)^{-1/2} \www_k(\tilde{\uuu}^*)\right],
\end{aligned}
\end{equation}
which implies that $\tilde{\uuu}^*$ is feasible to \eqref{eq:tight1} with $\beta=\beta_2$.
\end{lemma}

\begin{figure}
        \centering
        \includegraphics[width=.55\columnwidth]{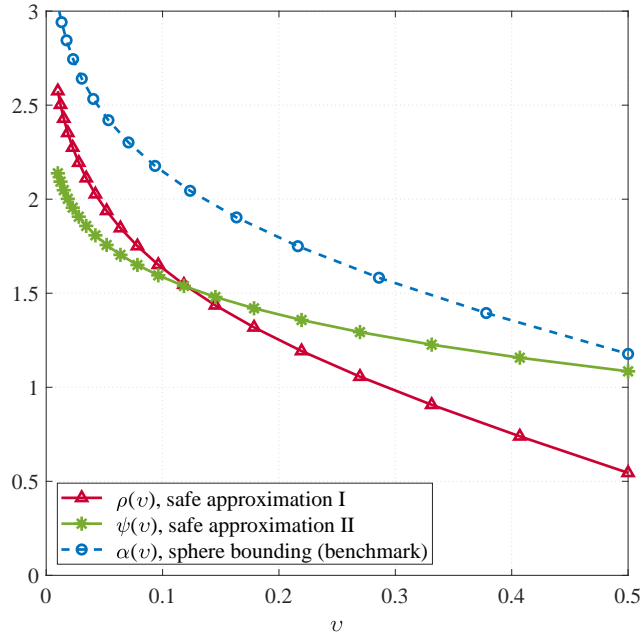}
        \caption{Plot of $\rho(\upsilon)$, $\alpha(\upsilon)$ and $\psi(\upsilon)$ as a function of the violation probability.}
        \label{fig:fac}
\end{figure}

It follows immediately from Lemma \ref{lem:3.5} and Lemma \ref{lem:4} that a relative comparison of the convex approximations $\mathrm{A}1$, $\mathrm{A}2$ and $\mathrm{B}$ boils down to just comparing $\rho(\upsilon)$, $\psi(\upsilon)$ and $\alpha(\upsilon)$. These three functions, however, depend on the violation probability $\upsilon$, as depicted in Fig. \ref{fig:fac} for $\upsilon\in(0,1/2]$. It can be observed from Fig. \ref{fig:fac} that for small values of $\upsilon$ below $\sim0.12$, which is of high practical interest, we have $\psi(\upsilon) \leq \rho(\upsilon) \leq \alpha(\upsilon)$. This means that a feasible solution to $\mathrm{B}$ is also feasible for $\mathrm{A}1$ and $\mathrm{A}2$, i.e., the optimal transmit power $\|\tilde{\uuu}^*\|^2$ obtained from $\mathrm{A}1$ and $\mathrm{A}2$ is no larger than that obtained from $\mathrm{B}$. Therefore, the robust convex approximations $\mathrm{A}1$ and $\mathrm{A}2$ are tighter (hence less conservative) than our benchmark $\mathrm{B}$. In a more precise order,
\begin{equation}\label{eq:tight3}
\mathcal{F}_{\mathrm{B}} \subseteq \mathcal{F}_{\mathrm{A}1} \subseteq \mathcal{F}_{\mathrm{A}2},
\end{equation}
where $\mathcal{F}_{(\cdot)}$ denotes the feasible set. It also follows from  \eqref{eq:tight3} that $\mathrm{A}2$ is tighter than $\mathrm{A}1$  in this range of $\upsilon$, i.e., under strict robustness settings. On the other hand, for higher values of $\upsilon$ up to $1/2$, which can be regarded as relaxed robustness conditions (but of course might be of less importance in a real system), we have $\rho(\upsilon) \leq \psi(\upsilon) \leq \alpha(\upsilon)$. This implies that $\mathrm{A}1$ provides a tighter convex approximation than $\mathrm{A}2$ in the high violation probability regime, but still $\mathrm{A}2$ is tighter than the benchmark approximation.

 \begin{table*}
        \caption{Complexity comparison of the non-robust and the proposed robust design approaches.}
        \label{tab:co}
        \centering
        \renewcommand{\arraystretch}{1}
        \begin{tabular}{ccc}
                \toprule
                Design problem & Complexity order $\left[\times \ln(\frac{1}{\epsilon})\right]$ & Dominating term [as $N,K\rightarrow\infty$]\\
                \midrule
                $\mathcal{P}1$ & \quad$2\sqrt{2K+2}\,.\,\mathcal{O}\left((2N+1)^3+(2K+1)(2N+1)(N+1)\right)$ & $\sqrt{K}\,.\,\mathcal{O}\left(N^3\right)\ln(\frac{1}{\epsilon})$
                \\
                \midrule
                $\mathcal{P}2$ & \quad$2\sqrt{4K+3}\,.\,\mathcal{O}\left((2N+1)^3+4KN^2(2N+1)+(2N+1)(N+1)\right)$ & $K\sqrt{K}\,.\,\mathcal{O}\left(N^3\right)\ln(\frac{1}{\epsilon})$
                \\
                \bottomrule
        \end{tabular}
\end{table*}


\section{Robust SINR-Constrained Power Minimization}\label{sec:optprob}

In this section, we aim to use the proposed robust approaches for the CI constraint obtained in the previous section in order to cast robust design formulations for the symbol-level precoder. We are particularly interested in an SINR-constrained power minimization problem which can expressed, in the non-robust form, by
\begin{equation}\label{eq:pm}
\begin{aligned}
\mathcal{P}1: \quad \underset{\tilde{\uuu}}{\mathrm{minimize}} & \quad \tilde{\uuu}^T\tilde{\uuu}\\
\mathrm{s.t.} & \quad \A_k \HHH_k \tilde{\uuu} \geq \sigma_k\sqrt{\gamma_k} \, \A_k \s_k, \; k=1,...,K,
\end{aligned}
\end{equation}
This design formulation aims at minimizing the total transmit power at each symbol time subject to CI constraints and given target SINRs $\gamma_k$ for all the users $k=1,...,K$. By introducing a slack variable $p\geq0$, it is further possible to recast \eqref{eq:pm} as
\begin{equation}\label{eq:pm2}
\begin{aligned}
\mathcal{P}1: \enspace \underset{\tilde{\uuu},p\geq0}{\mathrm{minimize}} & \quad p\\
\mathrm{s.t.} & \quad \A_k \HHH_k \tilde{\uuu} \geq \sigma_k\sqrt{\gamma_k} \, \A_k \s_k, \; k=1,...,K,\\
& \quad \tilde{\uuu}^T\tilde{\uuu} \leq p,
\end{aligned}
\end{equation}
which is more convenient for a later use in this section.
In the presentation of the design problem $\mathcal{P}1$ in \eqref{eq:pm}, it is assumed that all the users' channels are perfectly known to the transmitter. However, in the absence of such a knowledge, the design objectives and constraints are no longer guaranteed. For example, in addition to the error-iduced distortion of the CI regions at the users' receivers, the users may not be provided with the minimum required SINRs given by the target thresholds $\gamma_k,k=1,...,K$. Therefore, relying on robust formulations for the precoding design problem is essential in order to ensure the minimum SINR requirement of the users in any realizable case of the partially-known CSI. 

The robust counterpart of $\mathcal{P}1$ can be simply expressed by replacing the actual CI constraint with the worst-case robust constraint $\mathrm{W}$, in the case of spherical uncertainty, and either of the approximate constraints $\mathrm{A}1$, $\mathrm{A}2$, or $\mathrm{B}$ in the case of stochastic uncertainty. The resulting worst-case/stochastic robust formulation is then obtained as
\begin{equation}\label{eq:pmrob}
\begin{aligned}
\mathcal{P}2: \enspace \underset{\tilde{\uuu},p\geq0}{\mathrm{minimize}} & \quad p\\
\mathrm{s.t.} & \quad \text{either}\; \mathrm{W}, \mathrm{A}1, \mathrm{A}2, \; \text{or} \; \mathrm{B}, \; k=1,...,K,\\
& \quad \tilde{\uuu}^T\tilde{\uuu} \leq p.
\end{aligned}
\end{equation}
As summarized in Table \ref{tab:c}, the robust constraints $\mathrm{W}$, $\mathrm{A}1$, $\mathrm{A}2$, and $\mathrm{B}$ can all be formulated as second-order cone constraints, therefore the robust optimization problem $\mathcal{P}2$ falls within the class of convex conic quadratic programming (CQP). Notice, however, that while the non-robust formulation $\mathcal{P}1$ is always feasible, its robust counterpart $\mathcal{P}2$ may not share this property, as typical in robust optimization.

 
\subsubsection*{Computational Complexity Analysis}

We evaluate the computational complexity of the proposed robust design formulations based on the worst-case complexity analysis provided in \cite{complexity}, and compare the results with those of the original non-robust formulation. All the robust formulations, including worst-case and stochastic, are presented as CQPs, which can efficiently be solved via interior-point methods. In general, the arithmetic complexity of a generic interior-point method entails the Newton complexity as well as per-iteration computation cost. The Newton complexity basically refers to the number of steps required to reduce the duality gap by a constant factor, while the per-iteration complexity involves finding a new search direction at each step, and is subsequently dominated by the computation effort to assemble and solve a linear system of equations. In particular, we briefly overview the complexity bound for a CQP given in a generic form containing linear and (conic) quadratic constraints, to reach an $\epsilon$-solution (i.e., an $\epsilon$-optimal feasible solution) via a generic interior-point method.

For the conic quadratic program
\begin{equation}\label{eq:cqp}
\begin{aligned}
\underset{\x}{\mathrm{minimize}} & \quad \ccc_0^T\x\\
\mathrm{s.t.} & \quad \|\FFF_i\x+\bbb_i\| \leq \fff_i^T\x+g_i, \; i=1,...,m,\\
\mathrm{s.t.} & \quad \ccc_j^T\x \leq d_j, \; j=1,...,l,\\
& \quad \|\x\| \leq d_0,
\end{aligned}
\end{equation}
where $\FFF_i\in\mathbb{R}^{n_i\times n}, \bbb_i\in\mathbb{R}^{n_i}, \fff_i\in\mathbb{R}^{n},g_i\in\mathbb{R}$ for all $i=0,1,...,m$, and $\ccc_j\in\mathbb{R}^{n}, d_j\in\mathbb{R}$ for $j=0,1,...,l$, the complexity bound of an $\epsilon$-solution is of order
\begin{equation}\label{eq:lin2}
\mathcal{C}(\mathcal{P},\epsilon) = n\,\sqrt{l+2m}\left(n^2+l(n+1)+\sum_{i=1}^m n^2_i\right)\mathcal{O}(1).
\end{equation}

In the CQP formulation \eqref{eq:cqp}, $n$ can be read as the total number of optimization variables, and $n_i$ determines the size of the $i$th cone constraint, which is related to the dimension of the $i$th second-order cone, for all $i=1,...,m$. Notice that this generic form of CQP encompasses also the non-robust formulation in \eqref{eq:pm2}. Based on the above analysis, we are able to analyze the complexity of the robust CQP design formulation \eqref{eq:pmrob}, and compare it to that of its non-robust counterpart in \eqref{eq:pm2}. We also remark that
\begin{itemize}
        \item[i.] There are two real-valued second-order cone constraints associated with each user.
        \item[ii.] The slack variables $p$ in \eqref{eq:pmrob} can be merged into the vector $\tilde{\uuu}$, increasing the $i$th cone's dimension by one for all $i=1,...,m$.
\end{itemize}
Accordingly, for all design problems, the number of variables is equal to $2N+1$. The non-robust formulation \eqref{eq:pm2} has $2K+1$ linear inequalities plus one cone constraint of size $2N+1$, while the robust design formulation \eqref{eq:pmrob} involves $2K$ conic constraints of size $2N$ and one conic constraint of size $2N+1$ which corresponds to the power constraint. In Table \ref{tab:co}, the final complexity results obtained from \eqref{eq:cqp} are reported, where the dominating terms represent the largest complexity growth rate as $N,K\rightarrow\infty$ under the assumption $K\leq N$. It follows from Table \ref{tab:co} that for both design problems, the proposed robust formulations increase the computational complexity of precoding design by an order of $\mathcal{O}(K)$, compared to those of their non-robust counterparts.

 

\section{Simulation Results}\label{sec:sim}

\begin{figure}
	\centering
	\includegraphics[width=.55\columnwidth]{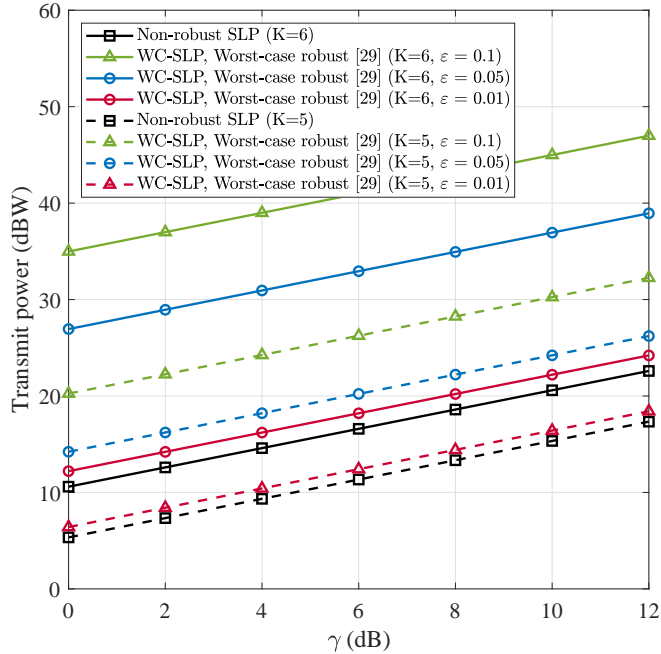}
	\caption{Average transmit power of the non-robust and the worst-case robust SLP schemes versus SINR target in a system with $N=6$.}
	\label{fig:p_wc}
\end{figure}

In this section, we present our simulation results to evaluate the performance of the proposed robust symbol-level precoding (SLP) schemes, and further to validate the analytic discussions provided in earlier sections. The optimization problems have been solved through MATLAB software by using CVX convex optimization package \cite{cvx}, and SeDuMi solver \cite{sedumi}. The following setup is adopted in all the simulation scenarios. We consider a downlink multiuser MISO system, employing an 8-ary phase-shift keying (8-PSK) modulation scheme. For all the users $k=1,...,K$, we set unit noise variances $\sigma_k^2=1$ and equal SINR requirements $\gamma_k=\gamma$. The estimate channel vectors $\hat{\h}_k,k=1,...,K$ are randomly generated according to the zero-mean circularly symmetric complex Gaussian distribution with unit variance, where the channel vectors of different users are independent, i.e., $\EXP\{\hat{\h}_k^H\hat{\h}_j\}=\OOO, \forall k,j=1,...,K,k\neq j$. We assume identical uncertainty regions for all the users' channels, i.e., $\varepsilon_k=\varepsilon,k=1,...,K$, in the case of spherical uncertainty region, and $\xi_k^2=\xi^2,k=1,...,K$, under stochastic uncertainty.

In Fig. \ref{fig:p_wc}, the transmit power performance of the proposed worst-case robust SLP (WC-SLP) is displayed versus SINR target $\gamma$ under the spherical uncertainty region with three different radii $0.01$, $0.05$ and $0.1$. As it might be expected, for larger uncertainty regions, higher transmission powers is needed in order to guarantee the system/users' requirements in case of any possible realization of the bounded CSI error. Furthermore, the performance results are depicted for two system dimensions with $N=K=6$, and $N=6$ and $K=5$. It follows from Fig. \ref{fig:p_wc} that the system requires less additional power to provide robustness to bounded CSI uncertainty for fewer number of users. For instance, in the case with $\varepsilon=0.01$, decreasing the number of users by one results in a reduction of around $6$ dBW in the average transmit power of the worst-case robust SLP.
We highlight that, for PSK modulations, the WC-SLP scheme shows the same performance as that of the worst-case robust symbol-level design in \cite{slp_chr}. However, as mentioned earlier, the method in \cite{slp_chr} is formulated only for constant envelope modulation schemes, whereas our proposed worst-case method does not have such a restriction and applies to a broader group of modulations. 

\begin{figure}[t]
        \begin{subfigmatrix}{2}
                \subfigure[]{\includegraphics[trim={.4in .4in .4in .4in},clip,width=.49\columnwidth]{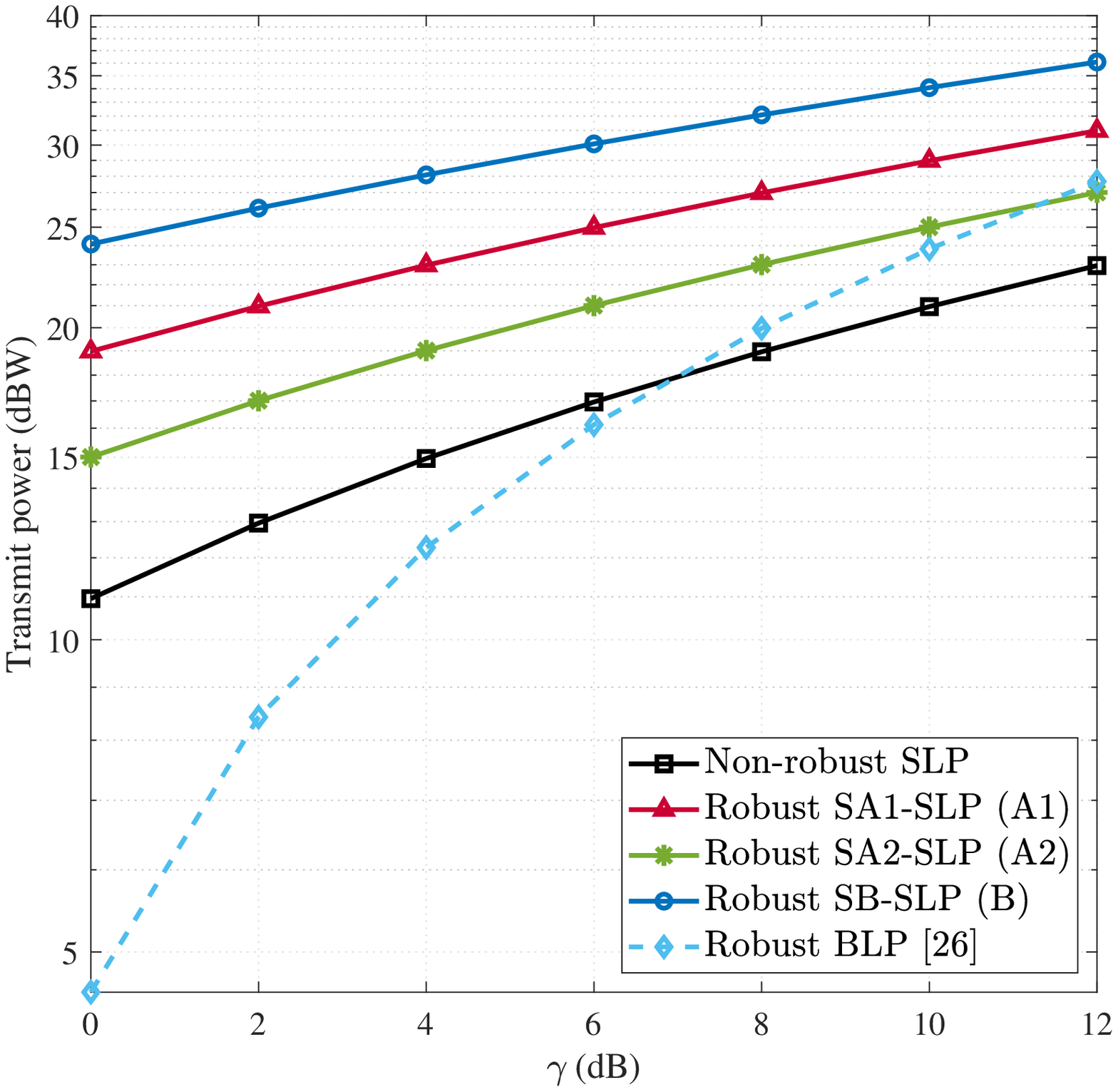}}
                \subfigure[]{\includegraphics[trim={.4in .4in .4in .4in},clip,width=.49\columnwidth]{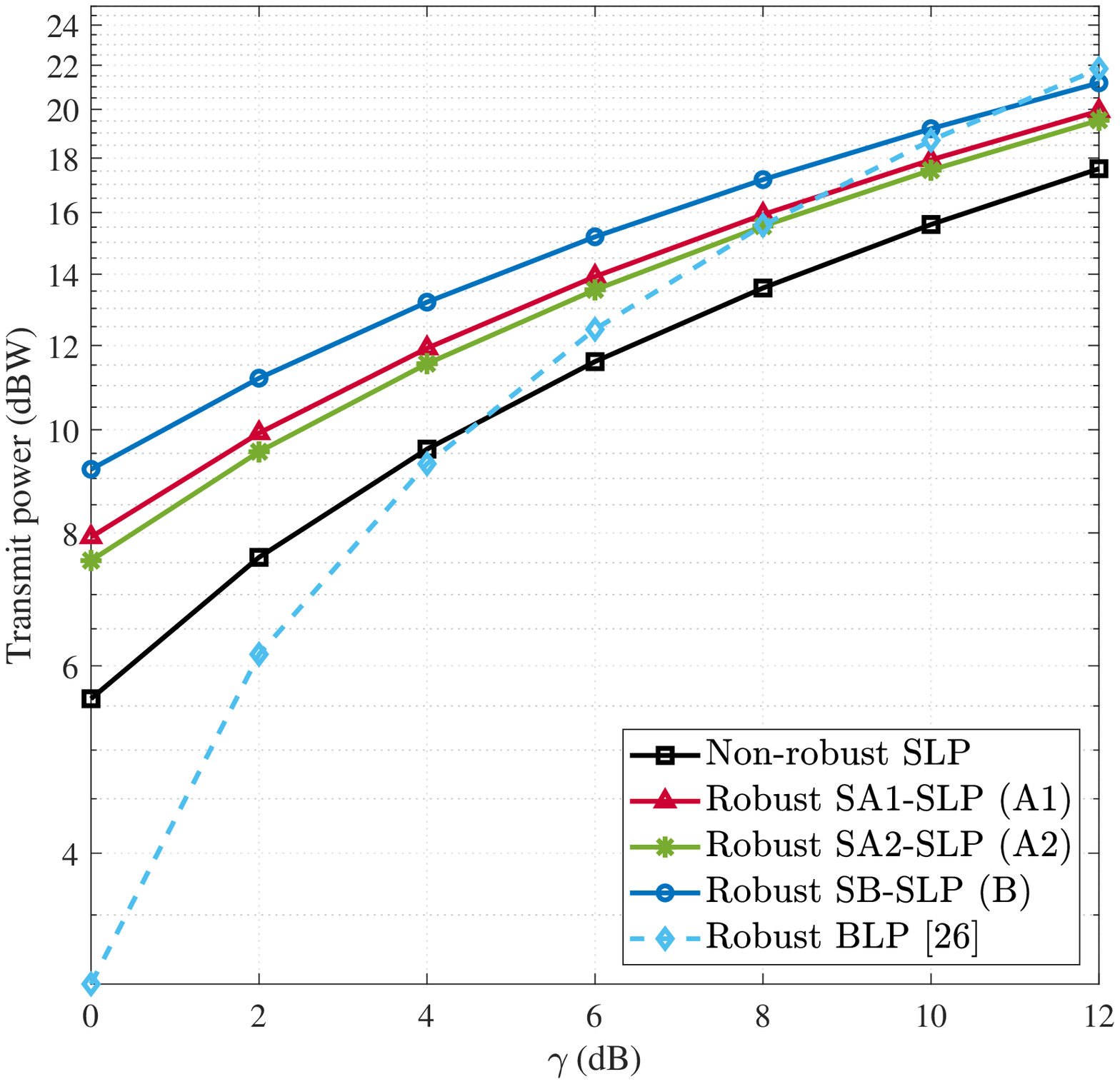}}
        \end{subfigmatrix}
        \caption{Average transmit power of different precoding schemes versus SINR target in a system under stochastic uncertainty with $\xi^2=0.004$ and $\upsilon=0.05$ (a) $N=K=6$ (b) $N=6$ and $K=5$.}
        \label{fig:p_54}
\end{figure}

\begin{figure}[t]
        \begin{subfigmatrix}{2}
                \subfigure[]{\includegraphics[trim={.4in .4in .4in .4in},clip,width=.49\columnwidth]{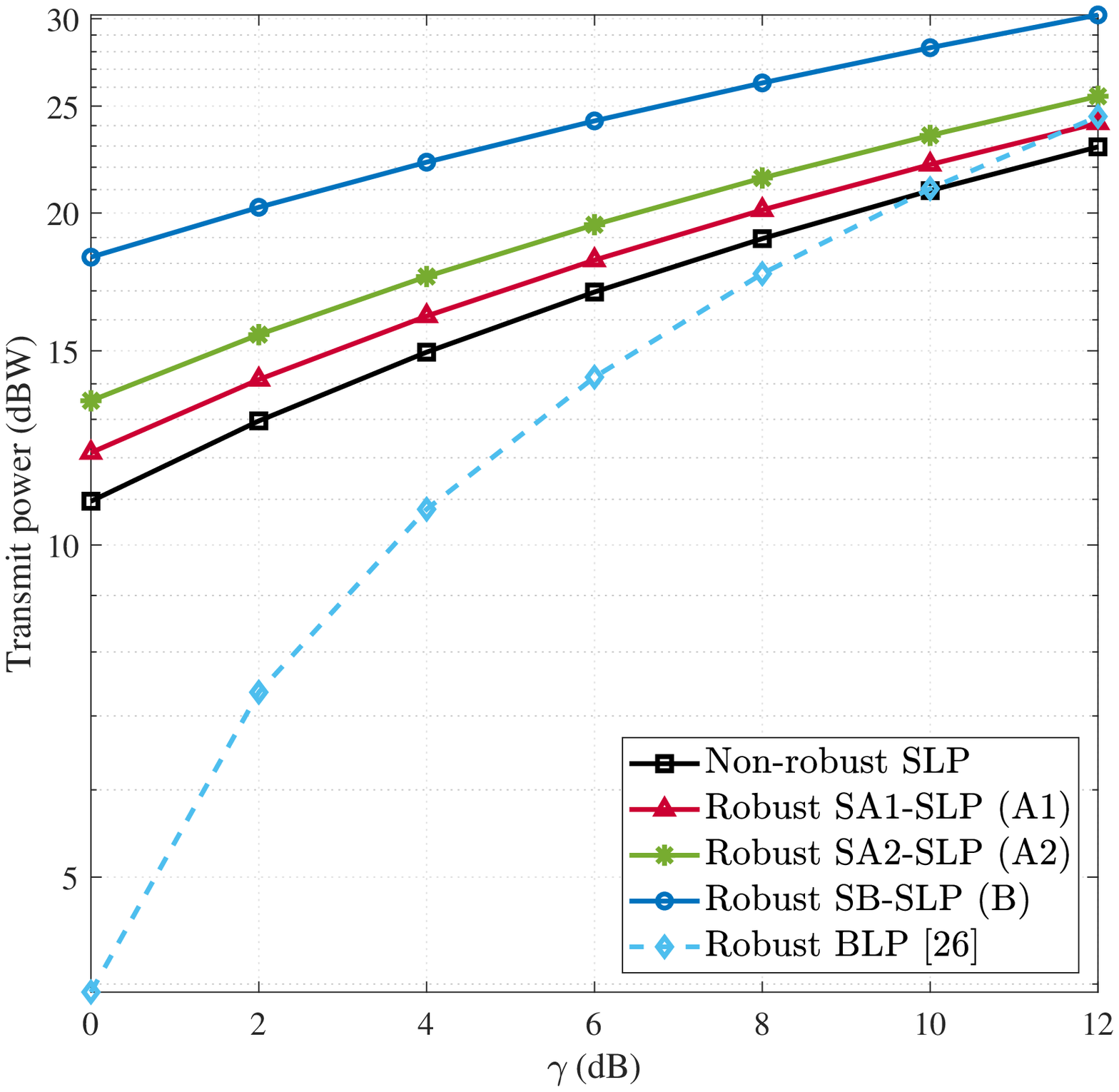}}
                \subfigure[]{\includegraphics[trim={.4in .4in .4in .4in},clip,width=.49\columnwidth]{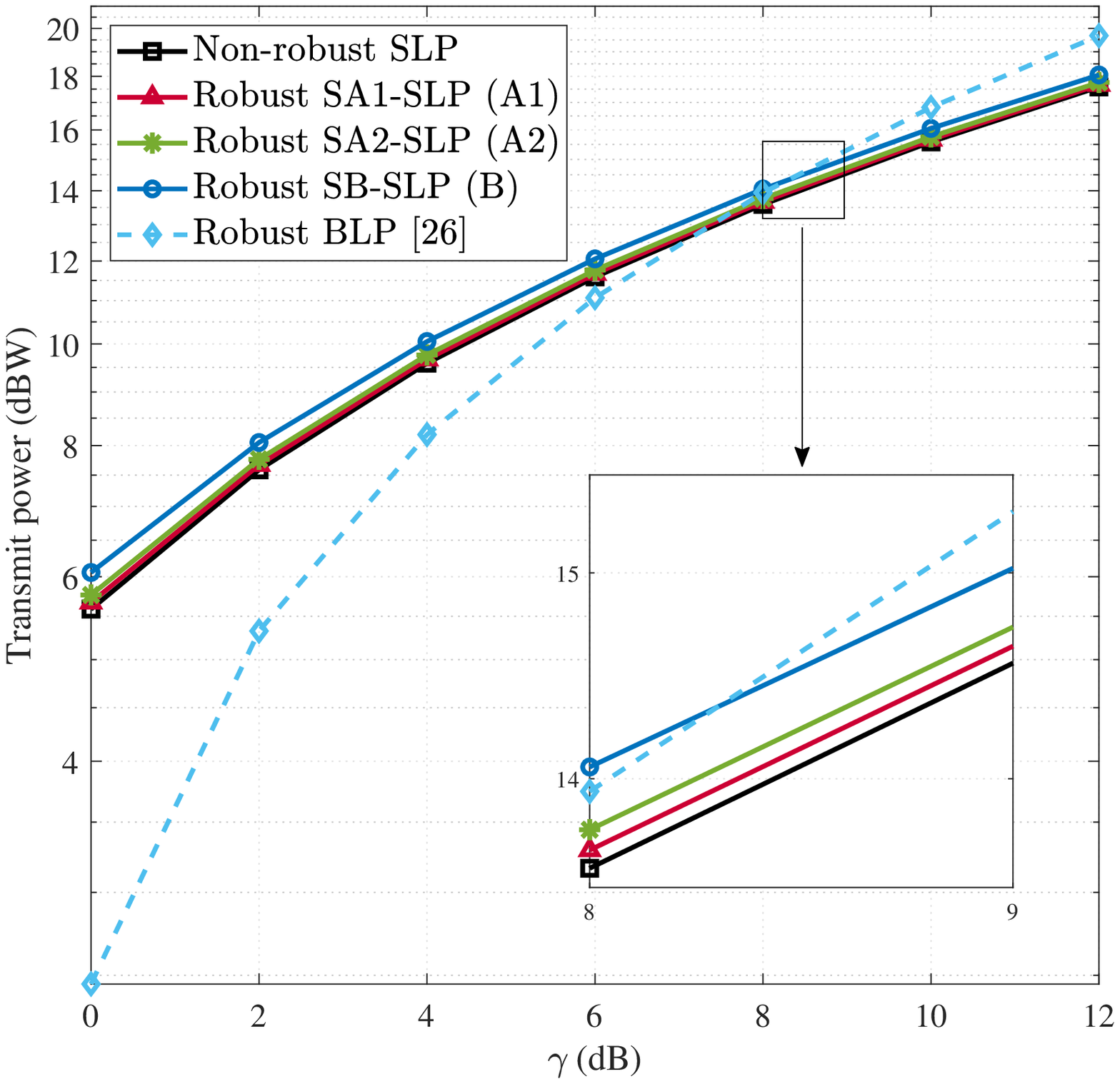}}
        \end{subfigmatrix}
        \caption{Average transmit power of different precoding schemes versus SINR target in a system under stochastic uncertainty with $\xi^2=0.001$ and $\upsilon=0.2$ (a) $N=K=6$ (b) $N=6$ and $K=5$.}
        \label{fig:p_21}
\end{figure}

Under the stochastically known CSI errors, we evaluate the performance of the downlink transmission in terms of the average consumed power versus SINR target obtained by different conventional and symbol-level precoding schemes. The simulation results are presented in Fig. \ref{fig:p_54} and Fig. \ref{fig:p_21}. The SLP approaches with robust CI constraints safe approximation I and II, and sphere bounding are respectively referred to as SA1-SLP, SA2-SLP and SB-SLP. We also show the results for a conventional (block-level) robust precoding scheme proposed in \cite{outage_conic}, labeled as robust BLP, which uses the Bernstein-type inequality to bound the outage probability of a given target rate $R$ (the target rate is connected to the SINR requirement via $\gamma=2^R-1$). Two stochastic uncertainty scenarios are investigated, each with an appropriate robustness consideration. The first scenario assumes a severe channel uncertainty with $\xi^2=0.005$, but imposes strict robust condition $\upsilon=0.05$ (which promises the service availability to the users in at least $95\%$ of times). In a second more relaxed scenario, a milder uncertainty with $\xi^2=0.001$ is assumed and the robust condition is set to be $\upsilon=0.2$. A common observation from Fig. \ref{fig:p_54} and Fig. \ref{fig:p_21} is that for an underloaded system with $K<N$, we have a larger feasible region brought by fewer number of robust CI constraints, and hence more degrees of freedom, to achieve lower transmit powers. It can be further observed that the performances of the proposed robust methods are always superior to those of the benchmark scheme SB-SLP (in both scenarios), as suggested by our tightness analysis. The results of the first scenario are shown in Fig. \ref{fig:p_54} for two different system dimensions. It has been verified that SA2-SLP provides robustness with a lower level of conservatism, hence a lower transmit power, whenever strict robust conditions are set for the system. In comparison with the SLP methods, the robust BLP scheme shows a better performance for low SINR targets, however it becomes more conservative as $\gamma$ increases. The SA1-SLP and SA2-SLP methods outperform the robust BLP scheme for $\gamma \geq 11$ dB and $\gamma \geq 8$ dB in a downlink system, respectively, with $K=6$ and $K=5$ users. This may suggest that the threshold on $\gamma$ (above which SLP performs better) reduces by decreasing the number of users. Nevertheless, as we will see later, the smaller transmission power of the robust BLP in the low SINR regime comes with a noticeably degraded symbol error rate performance. On the other hand, under relaxed robustness settings, it follows from Fig. \ref{fig:p_21} that the extra power needed for a robust transmission becomes smaller, or even insignificant particularly for the robust SLP methods with $\upsilon=0.2$; see Fig. \ref{fig:p_21} (b). Furthermore, it can be seen that the SA1-SLP method offers a less conservative robust scheme compared to SA2-SLP, in relaxed robust settings.

\begin{figure}
        \centering
        \includegraphics[width=.55\columnwidth]{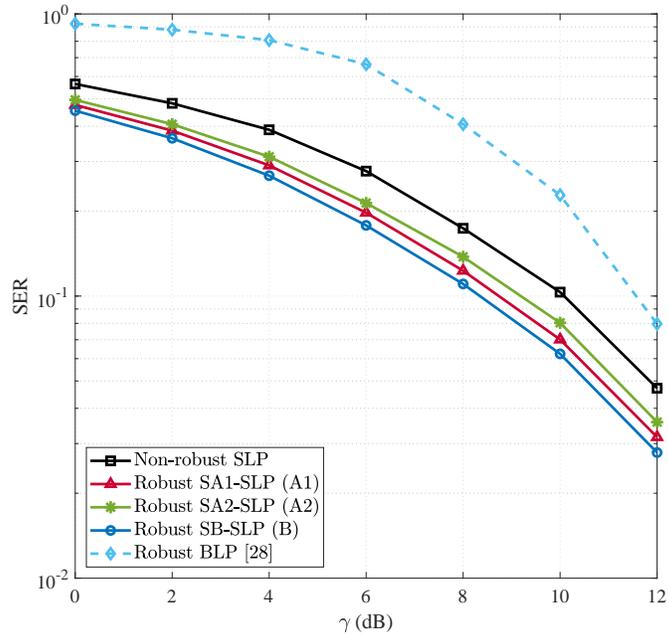}
        \caption{Average users' symbol error rate versus SINR target in a system with $N=K=6$, $\xi^2=0.004$, and $\upsilon=0.05$.}
        \label{fig:ser}
\end{figure}

The average users' symbol error rate (SER) for an uncoded transmission is shown in Fig. \ref{fig:ser} as a function of the SINR requirement $\gamma$, for different stochastic robust schemes. It can be observed that the robust BLP scheme has a higher SER than those of the SLP methods, though consuming less power in the depicted range of $\gamma$. However, the lower SER of the robust SLP methods is mostly an advantage of introducing the CI constraints in the precoder optimization problem. It can be also inferred from Fig. \ref{fig:ser} that a more conservative robust CI constraint provides lower SERs, but on the other hand leads to higher power consumptions.
This, however, means that the users are provided with higher SINRs than the required QoS level (i.e., $\gamma$), which may not be efficient in general, especially when the goal is to optimize the transmit power under a given SER target. In systems without such SER requirement, there is a power-performance tradeoff to be balanced, according to which the most efficient robust transmission scheme is preferred.

\begin{figure}
        \centering
        \includegraphics[width=.55\columnwidth]{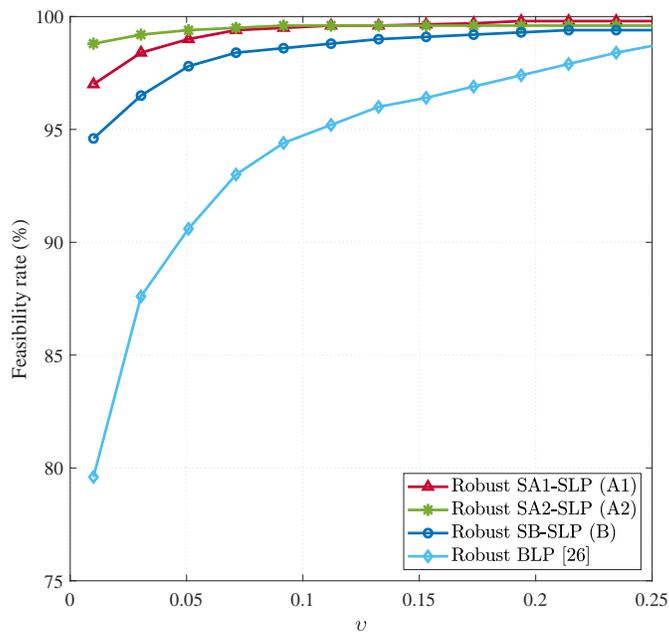}
        \caption{Feasibility rate over 2000 channel realizations as a function of the violation probability with $N=K=6$, $\gamma=5$ dB, and $\xi^2=0.004$.}
        \label{fig:feas_ep}
\end{figure}

\begin{figure}
        \centering
        \includegraphics[width=.55\columnwidth]{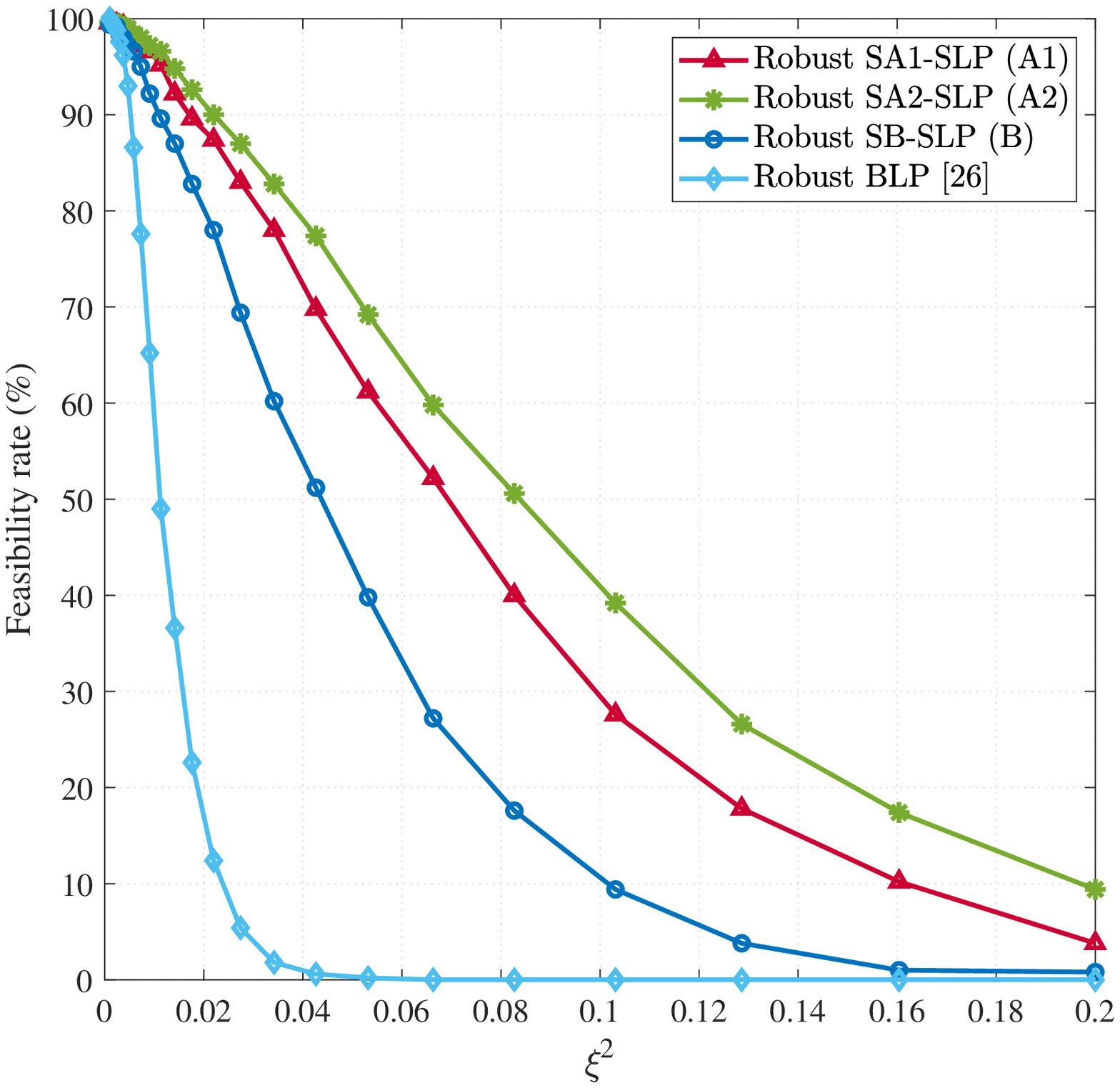}
        \caption{Feasibility rate over 2000 channel realizations as a function of the uncertainty variance with $N=K=6$, $\gamma=5$ dB, and $\upsilon=0.05$.}
        \label{fig:feas_var}
\end{figure}

In Fig. \ref{fig:feas_ep} and Fig. \ref{fig:feas_var}, the feasibility of different stochastic robust schemes is investigated with respect to the violation probability $\upsilon$ and the uncertainty variance $\xi^2$, respectively. For the sake of distinction, the results are presented only in the interval $\upsilon\in(0,0.25]$, however, based on our observations, the feasibility plot of each method shows an exact same behavior for violation probabilities up to $0.5$. As shown in Fig. \ref{fig:feas_ep}, both SA1-SLP and SA2-SLP outperform our benchmark SB-SLP in terms of feasibility. Furthermore, SA1-SLP and SA2-SLP are feasible more than $97\%$ of times in the whole range of $\upsilon$, while the robust BLP achieves this feasibility rate for violation probabilities higher than $0.15$. It is also worth noting that the feasibility rates of robust SLP methods in Fig. \ref{fig:feas_ep} validate our tightness analysis in Section \ref{sec:slp}, where we mentioned that the tighter the convex approximation is, the larger the feasible region will be. For example, according to Fig. \ref{fig:fac}, the probability bounding method becomes tighter than the safe approximation for $\upsilon>0.12$. This is verified by Fig. \ref{fig:feas_ep} in which the feasibility rate of SA1-SLP overtakes that of SA2-SLP at around $\upsilon=0.12$. Moreover, in a robustness setting with $\upsilon=0.05$, it can be seen from Fig. \ref{fig:feas_var} that all the robust SLP methods are feasible with higher rates in a much wider range of $\upsilon$ compared to the robust BLP. The robust BLP optimization appears to be barely feasible for uncertainty variances larger than $0.05$, while the SA1-SLP and SA2-SLP methods show feasibility rates of, respectively, $64\%$ and $71\%$ at $\xi^2=0.05$.

\begin{table}
        \caption{Average Simulation runtime (in seconds).}
        \label{tab:srt}
        \centering
        \renewcommand{\arraystretch}{.5}
        \begin{tabular}{lcccc}
                \toprule
                Precoding scheme & \multicolumn{4}{c}{Number of users $(K=N)$}\\
                \cmidrule{2-5}
                & $K=2$ & $K=4$ & $K=6$ & $K=8$\\
                \midrule
                Non-robust & $0.647$ & $0.652$ & $0.660$ & $0.702$\\
                \midrule
                Robust & $0.703$ & $0.759$ & $0.827$ & $0.931$\\
                \bottomrule
        \end{tabular}
\end{table}

Finally, in Table \ref{tab:srt}, we compare the simulation runtime of the non-robust and the robust SLP methods (either SA1-SLP or SA2-SLP) for different number of users, where the computation times are obtained by a relevant function of CVX. The results indicate that the robustness of symbol-level precoder is achieved with the price of an increased computation time, which coincides with the computational complexity discussion in Section \ref{sec:optprob}. More specifically, by increasing the number of users $K$, the runtime of the robust SLP optimization grows faster with respect to that of the non-robust scheme. In order to have a fair comparison of the results presented in this section, it should be also mentioned that the SLP approaches are typically more computationally demanding than the conventional block-level precoding schemes due to the required symbol-level processing.

\section{Concluding Remarks}\label{sec:con}

We addressed the (optimization) problem of a symbol-level precoded transmission scheme in a downlink MU-MIMO system under imperfect bounded or stochastic CSI knowledge at the transmitter. We formulated an optimization criterion aiming at minimizing the total transmit power subject to CI constraints as well as given QoS requirements in terms of the users' individual SINR targets. We developed robust CI constraints for each CSI uncertainty scenario and provided robust design formulations for the precoding optimization problem. With norm-bounded CSI errors, the worst-case robust formulation is obtained based on the conservation of guaranteeing the users' requirements for every possible realization of the channel within the uncertainty region. Under stochastic CSI uncertainty, a probabilistic approach is adopted to represent the optimization constraints, but led us to intractable expressions. We tackled this difficulty by deriving two computationally tractable approximate convex constraints with different levels of conservatism. A benchmark approximation was also derived based on the sphere bounding conservative method. Our analytical and simulation results indicate that both the proposed robust convex approximations outperform the benchmark, while each of which is superior to the other under different robust considerations. In comparison with conventional block-level robust schemes, although the proposed methods consume more power to achieve robustness in the low SINR regime, smaller transmit powers are observed with increasing the SINR target. However, the key advantages of the proposed robust SLP methods are better SER performances, as well as higher feasibility rates for wider ranges of violation probability and uncertainty variance, where the latter provides more service availability to the users in a practical multiuser system with imperfect CSI. Furthermore, it is shown via complexity analysis that the robustness of the SLP design comes with an increased computational complexity, particularly by an order of $K$ in the limiting case.

\appendices
\section{Proof of equality ($\mathrm{b}$) in \eqref{eq:vkv}}\label{app:b}

First, let $\QQQ_k\!\triangleq\!\EXP\{\VEC(\EEE_k)\VEC(\EEE_k)^T\}$ denote the covariance matrix of $\VEC(\EEE_k)$ as given in \eqref{Edel}.
It follows that
\begin{equation}\label{eq:appa2}
\QQQ_k = \frac{1}{2} \, \xi_k^2 \begin{bmatrix} \I_{N}\otimes\I_2 & \I_N \otimes \J_2 \\
\I_N \otimes \J_2^T & \I_{N}\otimes\I_2
\end{bmatrix},
\end{equation}
where we have used the facts that $(\I_N \otimes \J_2)^T=\I_N \otimes \J_2^T$ and $\I_{2N}=\I_N \otimes \I_2$. Now, the desired equality to be proven can be written as
\begin{equation}\label{eq:appa3}
\begin{aligned}
(\tilde{\uuu}^T \otimes \A_k) \; \QQQ_k (\tilde{\uuu} \otimes \A_k^T) = \frac{1}{2} \, \xi_k^2 \, (\tilde{\uuu}^T \otimes \A_k)(\tilde{\uuu} \otimes \A_k^T),
\end{aligned}
\end{equation}
Using the property $(\tilde{\uuu}^T \otimes \A_k)(\tilde{\uuu} \otimes \A_k^T)=(\tilde{\uuu}^T\tilde{\uuu}) \otimes (\A_k \A_k^T)$, equivalently, it is desired that
\begin{equation}\label{eq:appa4}
\begin{aligned}
(\tilde{\uuu}^T \otimes \A_k) \; \QQQ_k (\tilde{\uuu} \otimes \A_k^T) = \frac{1}{2} \, \xi_k^2 \, \|\tilde{\uuu}\|^2 (\A_k \A_k^T),
\end{aligned}
\end{equation}
We proceed by focusing on the left-hand side of \eqref{eq:appa4}. Let us denote $(\tilde{\uuu}^T \otimes \A_k) \QQQ_k (\tilde{\uuu} \otimes \A_k^T)\triangleq\GGG=[g_{ij}]_{2\times2}$ and $\tilde{\uuu}^T=[\uuu_\mathrm{R}^T,\uuu_\mathrm{I}^T]$, where $\uuu_\mathrm{R}=\mathrm{Re}(\uuu)$ and $\uuu_\mathrm{I}=\mathrm{Im}(\uuu)$. Thus, considering $\A_k=[\aaa_{k,1},\aaa_{k,2}]^T$, we have
\begin{equation}\label{eq:appa5}
\begin{aligned}
\GGG = \frac{1}{2} \, \xi_k^2 \begin{bmatrix} \uuu_\mathrm{R}^T\otimes\aaa_{k,1}^T & \uuu_\mathrm{I} \otimes \aaa_{k,1}^T\\
\uuu_\mathrm{R}^T \otimes \aaa_{k,2}^T & \uuu_\mathrm{I} \otimes \aaa_{k,2}^T
\end{bmatrix} \times \begin{bmatrix} \I_{N}\otimes\I_2 & \I_N \otimes \J_2 \\
\I_N \otimes \J_2^T & \I_{N}\otimes\I_2
\end{bmatrix}\times \begin{bmatrix} \uuu_\mathrm{R} \otimes \aaa_{k,1} & \uuu_\mathrm{R} \otimes \aaa_{k,2} \\
\uuu_\mathrm{I} \otimes \aaa_{k,1} & \uuu_\mathrm{I} \otimes \aaa_{k,2}
\end{bmatrix}.
\end{aligned}
\end{equation}
Foe the sake of simplicity, the term $\frac{1}{2} \, \xi_k^2$ is omitted from the next equation, but it will appear in the final derivation. 
The matrix multiplication in the right-hand side of \eqref{eq:appa5} can be evaluated and simplified as
\begin{subequations}\label{eq:appa7}
        \begin{align}
        g_{11} &= \left(\uuu_\mathrm{R}^T\uuu_\mathrm{R}\!+\!\uuu_\mathrm{I}^T\uuu_\mathrm{I}\right)\aaa_{k,1}^T\aaa_{k,1}\!+\!2\,\uuu_\mathrm{R}^T\uuu_\mathrm{I}\otimes \aaa_{k,1}^T\J_2\aaa_{k,1}, \label{eq:g11}\\
        g_{12} &= g_{21} = \left(\uuu_\mathrm{R}^T\uuu_\mathrm{R}+\uuu_\mathrm{I}^T\uuu_\mathrm{I}\right)\aaa_{k,1}^T\aaa_{k,2} + 2\,\uuu_\mathrm{R}^T\uuu_\mathrm{I}\otimes \left(\aaa_{k,1}^T\J_2\aaa_{k,2}
        +\aaa_{k,1}^T\J_2^T\aaa_{k,2}\right), \label{eq:g12}\\
                g_{22} &= \left(\uuu_\mathrm{R}^T\uuu_\mathrm{R}\!+\!\uuu_\mathrm{I}^T\uuu_\mathrm{I}\right)\aaa_{k,2}^T\aaa_{k,2}\!+\!2\,\uuu_\mathrm{R}^T\uuu_\mathrm{I}\otimes \aaa_{k,2}^T\J_2\aaa_{k,2}, \label{eq:g22}
        \end{align}
\end{subequations}
where in simplifications, we have frequently used the fact that
$(\XXX \otimes \YYY)(\WWW \otimes \ZZZ)=(\XXX \WWW \otimes \YYY \ZZZ)$, for any given matrices $\XXX,\YYY,\WWW,\ZZZ$ with appropriate dimensions. It is easy to verify that $\aaa_{k,1}^T\J_2\aaa_{k,1}=\aaa_{k,1}^T\J_2^T\aaa_{k,1}=\OOO$, and further $\aaa_{k,1}^T\J_2\aaa_{k,2}
+\aaa_{k,1}^T\J_2^T\aaa_{k,2}=\aaa_{k,1}^T(\J_2+\J_2^T)\aaa_{k,2}=\OOO$. Moreover, it directly follows from the definition of $\tilde{\uuu}$ that $\uuu_\mathrm{R}^T\uuu_\mathrm{R}+\uuu_\mathrm{I}^T\uuu_\mathrm{I}=\tilde{\uuu}^T\tilde{\uuu}$. Applying all these notes to \eqref{eq:g11}-\eqref{eq:g22}, the entries of $\GGG$ are obtained as
\begin{subequations}\label{eq:appa8}
        \begin{align}
        g_{11} &= \|\tilde{\uuu}\|^2 \|\aaa_{k,1}\|^2, \label{eq:g113}\\
        g_{12} &= g_{21} = \|\tilde{\uuu}\|^2 \, \aaa_{k,1}^T\aaa_{k,2}, \label{eq:g123}\\
        g_{22} &= \|\tilde{\uuu}\|^2 \|\aaa_{k,2}\|^2. \label{eq:g113}
        \end{align}
\end{subequations}
Merging the results in \eqref{eq:appa8} yields
\begin{equation}\label{eq:appa10}
\begin{aligned}
\GGG = \frac{1}{2} \, \xi_k^2 \, \|\tilde{\uuu}\|^2 (\A_k \A_k^T),
\end{aligned}
\end{equation}
as required.

\section{Derivation of equivalent SOC formulation for $\mathrm{A}2$}\label{app:soc}

The derivation is essentially based on Lemma \ref{lem:1}. We denote
$$
\XXX \triangleq \begin{bmatrix}
-\frac{\bar{w}_{k,1}}{\psi(\upsilon)} & 0 \\
0 & -\frac{\bar{w}_{k,2}}{\psi(\upsilon)}
\end{bmatrix},\;\;
\YYY \triangleq \begin{bmatrix}
1 & 0 & 0 & 0 \\
0 & 0 & 0 & 1 \\
\end{bmatrix},
$$
$$
\ZZZ \triangleq \begin{bmatrix}
-\frac{\bar{w}_{k,1}}{\psi(\upsilon)} & 0 & 0 & 0 \\
0 & -\frac{\bar{w}_{k,2}}{\psi(\upsilon)} & 0 & 0 \\
0 & 0 & -\frac{\bar{w}_{k,1}}{\psi(\upsilon)} & 0\\
0 & 0 & 0 & -\frac{\bar{w}_{k,2}}{\psi(\upsilon)}
\end{bmatrix}.
$$
Accordingly, the constraint \eqref{eq:safe3} can be equivalently implied by the following two semidefinite restrictions:
\begin{subequations}\label{eq:appb1}
\begin{align}
 \XXX \succeq 0, \\
\ZZZ - \YYY^T\XXX^{-1}\YYY \succeq 0. \label{eq:appb12}
\end{align}
\end{subequations}
The second restriction in \eqref{eq:appb12}, after doing the matrix products and some simple algebra, can be written as
\begin{equation}\label{eq:appb2}
\setlength\arraycolsep{1pt}
\begin{bmatrix}
-\frac{\bar{w}_{k,1}}{\psi(\upsilon)} + \frac{\psi(\upsilon)}{\bar{w}_{k,1}} & 0 & 0 & 0 \\
0 & -\frac{\bar{w}_{k,1}}{\psi(\upsilon)} & 0 & 0 \\
0 & 0 & -\frac{\bar{w}_{k,2}}{\psi(\upsilon)} & 0 \\
0 & 0 & 0 & -\frac{\bar{w}_{k,2}}{\psi(\upsilon)} + \frac{\psi(\upsilon)}{\bar{w}_{k,2}} \\
\end{bmatrix} \succeq 0.
\end{equation}
from which it is clear that \eqref{eq:appb12} further implies the restriction $\XXX \succeq 0$, hence it is necessary and sufficient for \eqref{eq:safe3}. We then rearrange \eqref{eq:appb2} in a more convenient form and decompose it into two semidefinite constraints as
\begin{subequations}\label{eq:appb3}
\begin{align}
\frac{-1}{\psi(\upsilon)}\, \pmb{D}_{\bar{\www}_k} \succeq 0, \label{eq:appb32}\\
\frac{-1}{\psi(\upsilon)}\, \pmb{D}_{\bar{\www}_k} + \psi(\upsilon)\, \pmb{D}^{-1}_{\bar{\www}_k} \succeq 0, \label{eq:appb31}
\end{align}
\end{subequations}
with $\pmb{D}_{\bar{\www}_k}\triangleq\diag(\bar{\www}_k)$. It should be noticed that the restriction \eqref{eq:appb32} is in fact equivalent to $\pmb{D}_{\bar{\www}_k} \preceq 0$, which is also implied by the assumption $\upsilon\in(0,1/2]$; see Remark 1. Further, note that $\mathrm{erfc}(\cdot)$ is non-negative in the interval $(0,1]$, so is $\psi(\upsilon)$. Now, multiplying both sides of \eqref{eq:appb31} by $\pmb{D}_{\bar{\www}_k}$, and imposing the restriction \eqref{eq:appb32} which changes the direction of the inequality, both of the constraints \eqref{eq:appb31} and \eqref{eq:appb32} can be simultaneously expressed by
\begin{equation}\label{eq:appb5}
\frac{-1}{\psi(\upsilon)}\, \pmb{D}^2_{\bar{\www}_k} + \psi(\upsilon)\, \I \preceq 0.
\end{equation}
Since $\pmb{D}_{\bar{\www}_k} \preceq 0$ and diagonal, from \eqref{eq:appb5} by taking square root, we obtain
\begin{equation}\label{eq:appb6}
\frac{1}{\psi(\upsilon)}\, \pmb{D}_{\bar{\www}_k} + \I \preceq 0,
\end{equation}
which can be written in the vector form as
\begin{equation}\label{eq:appb7}
\frac{-1}{\psi(\upsilon)}\, \bar{\www}_k \geq \pmb{1}.
\end{equation}
Replacing $\bar{\www}_k$ with $(\sqrt{2}/\xi_k \|\tilde{\uuu}\|)(\A_k \A_k^T)^{-1/2} \www_k(\tilde{\uuu})$, it is then routine to show that \eqref{eq:appb7} is equivalent to
\begin{equation}\label{eq:appb8}
\|\tilde{\uuu}\| \, \pmb{1} \leq \frac{-\sqrt{2}}{\psi(\upsilon) \, \xi_k}(\A_k \A_k^T)^{-1/2} \www_k(\tilde{\uuu}),
\end{equation}

\section*{Acknowledgment}
The authors are supported by the Luxembourg \mbox{National} Research Fund (FNR) under CORE Junior project: C16/IS/11332341 Enhanced Signal Space opTImization for satellite comMunication Systems (ESSTIMS).


\end{document}